\documentclass[fleqn,usenatbib]{mnras}

\usepackage{newtxtext,newtxmath}
\usepackage[T1]{fontenc}
\usepackage{color}
\usepackage[usenames,dvipsnames]{xcolor}
\DeclareRobustCommand{\VAN}[3]{#2}
\let\VANthebibliography\thebibliography
\def\thebibliography{\DeclareRobustCommand{\VAN}[3]{##3}\VANthebibliography}

\usepackage{graphicx}	% Including figure files
\usepackage{amsmath}	% Advanced maths commands

\title[Pair Counting without Binning]{Pair Counting without Binning - A New Approach to Correlation Functions in Clustering Statistics}

\author[Shi Yue et.al.]
{Shiyu Yue$^{1}$,
Longlong Feng$^{1}$\thanks{E-mail: flonglong@mail.sysu.edu.cn},
Wenjie Ju$^{1}$,
Jun Pan$^{2,3}$,
Zhiqi Huang$^{1}$,
Feng Fang$^{1}$,
Zhuoyang Li$^{4}$,
\newauthor  
Yan-Chuan Cai$^{5}$,
Weishan Zhu$^{1}$
\\
$^{1}$School of Physics and Astronomy, Sun Yat-sen University, 2 Daxue Road, Tangjia, Zhuhai, 519082, China \\
$^{2}$Chinese Academy of Sciences South America Center for Astronomy, National Astronomical Observatories, CAS, Beijing, 100101, China \\
$^{3}$College of Earth Sciences, Guilin University of Technology, Guilin, 541004, China \\
$^{4}$Department of Astronomy, Tsinghua University, Beijing 100084, China \\
$^{5}$Institute for Astronomy, University of Edinburgh, Blackford Hill, Edinburgh, EH9 3HJ, UK \\
}

\date{Accepted XXX. Received YYY; in original form ZZZ}

\pubyear{2024}

\begin{document}
\label{firstpage}
\pagerange{\pageref{firstpage}--\pageref{lastpage}}
\maketitle

\begin{abstract}
 
This paper presents a novel perspective on correlation functions in the clustering analysis of the large-scale structure of the universe. We begin with the recognition that pair counting in bins of radial separation is equivalent to evaluating counts-in-cells (CIC), which can be modelled using a filtered density field with a binning-window function. This insight leads to an in situ expression for the two-point correlation function (2PCF). Essentially, the core idea underlying our method is to introduce a window function to define the binning scheme, enabling pair-counting without binning. This approach develops an idea of generalised 2PCF, which extends beyond conventional discrete pair counting by accommodating non-sharp-edged window functions. In the context of multiresolution analysis, we can implement a fast algorithm to estimate the generalised 2PCF. To extend this framework to N-point correlation functions (NPCF) using current optimal edge-corrected estimators, we developed a binning scheme that is independent of the specific parameterisation of polyhedral configurations. In particular, we demonstrate a fast algorithm for the three-point correlation function (3PCF), where triplet counting is accomplished by assigning either a spherical tophat or a Gaussian filter to each vertex of triangles. Additionally, we derive analytical expressions for the 3PCF using a multipole expansion in Legendre polynomials, accounting for filtered field (binning) corrections. Our method provides an exact solution for quantifying binning effects in practical measurements and offers a high-speed algorithm, enabling high-order clustering analysis in extremely large datasets from ongoing and upcoming surveys such as Euclid, LSST, and DESI.

\end{abstract}


%%%%%%%%%%%%%%%%%%%%%%%%%%%%%%%%%%%%%%%%%%%%%%%%%%

%%%%%%%%%%%%%%%%% BODY OF PAPER %%%%%%%%%%%%%%%%%%

\section{Introduction}

The large-scale inhomogeneities of cosmic density fields are seeded by primordial quantum fluctuations in the very early universe, which are predicted to be Gaussian distributed in the standard inflationary scenario \citep{guth1982, hawking1982}. Statistically, a Gaussian random field is fully characterised by its mean and second-order statistics, including the 2PCF and its Fourier counterpart, the power spectrum; these are the fundamental clustering indicators of the large-scale distribution of mass traced by galaxies \citep[e.g.][and references therein]{peebles1980, Hamilton1988, HawkinsEtal2003, Yang2003, Eisenstein_2005, Li2006, zehavi2011, Wechsler2018, DESI_BAO2024} and place strong constraints on the cosmological models \citep[e.g.][]{cole2005, Eisenstein2005, Tegmark2006, Blake2011, Beutler2011, Percival2011, Anderson2014, Hildebrandt2016, Alam2017, Ivanov2020, Alam2021_SDSS}. However, driven by gravitational instability, the subsequent growth of density perturbations leads to highly nonlinear clustering on large scales, displaying significant non-Gaussian features. These non-Gaussianities cause information originally encoded in 2-point statistics to leak and cascade into a hierarchy of correlation functions at higher orders. The lowest order of non-Gaussian statistics is the 3PCF (the Fourier transform of which is the bispectrum), which provides complementary information about the background cosmology \citep[e.g.][]{Gagrani2017, Agarwal2021, Alam2021_SDSS, Gualdi2021, Samushia2021, Novell-Masot2023} and an independent probe to models of structure formation \citep[e.g.][]{Jing1998, Gaztanaga1994, Frieman1994, Scoccimarro2001, Jing2004, Kayo2004, Gaztanaga2005, PanSzapudi2005b, Nichol2006, KulkarniEtal2007, Marín2008, Marín2011, McBrideEtal2011, MarinEtal2013, GilMar2015, GilMar2017, Guo2015, Slepian2017, Pearson2018, Slepian2018, Veropalumbo2021, Sugiyama2023}.

To place stringent constraints on cosmological parameters, the error budgets for modern clustering analysis turn out to be very tight, especially the 2PCF, of which per cent or even sub-percent level errors are generally demanded in theoretical modelling and estimation from samples. It is then pivotal to devise efficient and robust methods to estimate correlation functions with datasets generated by galaxy surveys and simulations. Understanding and controlling systematic biases and variances induced by sample limitations and estimation algorithms becomes critical. It has been known that cosmic biases and errors roughly originate from three categories of effects: discreteness, finite volume, and geometrical edges \citep[including various {\it masks}, ][]{SzapudiColombi1996}, a good estimator should be able to suppress them and their joint effects substantially, being unbiased and optimal to variances. For the 2PCF, the two estimators proposed by \citet{LS1993} and \citet{Hamilton1993}, respectively, are numerically superior to others \citep{KerscherEtal2000}. The estimators of \citet{SzapudiSzalay1998} as a generalisation of \citet{LS1993} are recommended for higher-order correlation functions. 

It is acknowledged that estimators developed by \citet{LS1993} and \citet{SzapudiSzalay1998} are optimal under the condition that correlations on large scales vanish and the data sets are voluminous and densely populated. Ongoing and upcoming large-scale structure surveys map tens of millions to billions of galaxies over wide sky areas at significant depths, e.g., Euclid \citep{Laureijs2011}, CSST \citep{Zhan2011}, Dark Energy Spectroscopic Instrument (DESI) survey \citep{DESI2016}, Nancy Grace Roman Space Telescope \citep{Akeson2019} and Vera Rubin Observatory \citep{Ivezic2019}, and are generally supported by numerous mock catalogues created from extensive N-body simulations. A large number of data points in large volumes significantly reduces the statistical uncertainties induced by sparse sampling and inadequate volume. Nevertheless, some previously overlooked biases inherent in standard estimators and their numerical implementations now seem non-negligible. Various schemes have been proposed to improve standard procedures, including using reference catalogues other than Poisson randoms \citep{keihanen2019estimating, DavilaEtal2021, Schulz2023}, providing precise pair counting of random catalogues \citep{breton2021fast, he2021fast, kerscher2022improving}, applying post-processes to optimise estimators by case \citep{vargas2013optimized, BaxterRozo2013, Tessore2018, SosaNiz2020, Fisher2021}, and reducing binning effects \citep{JangJi2017,Fisher2021}.

Meanwhile, a large number of data points presents a significant computational challenge in measuring 2/3 PCFs. Their estimators are generally built on the basis of pair/triplet counting. Unlike power spectrum and bispectrum, whose estimation can benefit from the efficient fast Fourier transform (FFT), for the $N$-th order correlation function, counting all $N$-tuplets of $N_p$ discrete points via a brute-force approach will scale as $N_p^N$. Consequently, $N$-tuplet counting becomes the primary computing process to which fast algorithms aim to speed up or transform \citep[e.g.][]{Moore2001, SlepianEisenstein2016, DeminaEtal2018, Donoso2019, Sinha2020, PhilcoxEtal2022, PhilcoxSlepian2022, BrownEtal2022, Sunseri2023, Zhao2023}. 

For a discrete point set, the probability of finding pairs separated exactly by the specified scale is infinitesimal; therefore, counting $N$-tuplet can only be executed in a specific sequence of bins parameterized by the geometric shape of $N$-point configurations, currently referred to as a binning scheme. Variants in the binning scheme may produce different biased measurements. However, for a given N-point clustering statistics, the physical information inferred from the measurements should be independent of the binning scheme, unless the biased measurements due to the binning effect can be corrected appropriately. For example, to measure 2PCF, a sequence of separated radial bins with finite widths centred on the interested scales must be constructed to count neighbours such that the measured 2PCF is effectively averaged over each bin. In practice, binning setup is a trade-off between bias and variance: small bin sizes reduce bias but increase variances, while large bins may smooth small-scale fluctuations while inevitably producing biased measurements. To address this binning issue, there are two different approaches: either to search for an optimal binning plan \citep[e.g.][]{PercivalEtal2014, JangJi2017} and incorporate the binning effect into their models \citep{Bailoni2017}, or to find a way to reconstruct a continuous field from the given sample and make an estimation without binning \citep{Feng2007,Fisher2021}. 

The early motivation of this paper is to develop a fast algorithm for NPCF integrated into the {\tt Hermes} toolkit (HypER-speed MultirEsolution cosmic Statistics) - an open source parallel/GPU-accelerated tool for cosmic statistics in Python (Feng et al., in preparation). The core algorithm implemented in {\tt Hermes} is updated from the {\tt MRACS} (Multi-Resolution Analysis for Cosmic Statistics) scheme\citep{Feng2007}, in which a spatial point process is approximated by a continuous distribution decomposed in terms of a set of compact basis functions, and the CIC distribution can be produced by convolution of the reconstructed density field with a low-pass window function specified by the given geometric volume. 

While revisiting the 2PCF calculation, it is crucial to recognise that pair counting in bins is fundamentally a CIC operation, which can be executed by convolving the original density field with a window function defined by the bins. In view of practical measurement, the 2PCF in bins acts as an in situ cross-correlation between the original density field and its filtered one. This insight bridges the theoretical model and practical binning measurement of 2PCF, and generalises the {\tt MRCAS} algorithm of \citet{Feng2007}, enabling an $\mathscr{O}(N_g\log N_g)$ ($N_g$: the number of grids) estimation of the 2PCF.  Additionally, this approach offers practical flexibility in using non-sharp-edged binning window functions, which can be tailored for specific scientific objectives. This method is also particularly well suited to calculate a series of conditional cumulants, which are angle-averaged correlation functions \citep{PanSzapudi2005a, PanSzapudi2005b}.

However, when extending to higher-order correlation functions, such as the $N$-point correlation function (NPCF) for $N > 2$, the challenge arises of determining which set of geometric parameters to include in the bins for a given polyhedral configuration. For instance, a commonly used binning scheme for measuring the 3PCF involves binning three parameters: two side lengths of a triangle and the angle between them (e.g. \citealt{Gaztanaga2005, Kulkarni2007, Marín2008, Marín2011}). Unlike the measurement of the 2PCF in real space, where the binning window function is spherically symmetric, this 3PCF binning scheme leads to a set of spatially non-uniform window functions in the CIC (bins) calculation.

To fully exploit the advantages of the {\tt MRACS} algorithm for fast cell-in-cell calculations, we adopt a specific strategy in designing the binning scheme. Since the binning window function assigned to each vertex of a triangle can, in principle, be independent of its shape parameters, we aim to maximise symmetry by ensuring that the binning window function is invariant under spatial translation and rotation, enabling a uniformly spatial binning in the triplet counting. With this strategy, the most straightforward choice for the binning window function is a spherical tophat with a radius corresponding to the bin width. This approach simplifies triplet counting in 3PCF measurement as triple-sphere CIC. 

Since CIC, binning, and filtering share the same mathematical operation, a binning scheme can be converted to a filtered operation on the original density field. Unlike traditional binning schemes, the filters can be independent of the parameterisation of the polyhedral configuration. Consequently, NPCFs of arbitrary spatial configurations can be efficiently estimated by filtered fields, i.e., by binned density fields \citep{Slepian2018}. In the sense of traditional counting, it is a counting without binning. The expense of dropping off binning in this way is that one needs to take the smoothing effects into account in theoretical modelling. We show that at least for 2/3PCFs of the filtered field, the model calculation is not more intimidating than the one accounting for binning effects. Compared with binned correlation functions of the discrete sample, correlation functions of the filtered field are particularly useful for investigating features at large scales. The cumulant correlators \citep{SzapudiSzalay1997, Szapudi1998, MunshiEtal2000} are actually special cases of the correlation functions of the filtered field. 

The paper is organised as follows. Section 2 introduces a novel in-situ concept of 2PCF and discusses how it can be applied to quantify the binning effect in the 2PCF calculation. Furthermore, a generalised definition of 2PCF is also presented. Section 3 outlines the {\tt MRACS} scheme for CIC statistics within the multiresolution analysis framework and its application to computing 2PCF. In Section 4, we perform the numerical tests of the generalised 2PCF against the publicly available simulation samples. Section 5 extends to 3PCF measurements using a triple-sphere binning scheme. We present the corresponding analytical expression of 3PCF, and the numerical tests. Finally, we summarise the paper and make the concluding remarks in Section 6. 
In addition, the convergence and performance tests for 3PCF are presented in the Appendix.  

\section{2PCF: an alternative view from ex-situ to in-situ}

\subsection{Quantifying the binning effect in 2PCF}
In a statistically homogeneous and isotropic random density field, the 2PCF is the Fourier counterpart of the power spectrum
\begin{equation}\label{eq:2PCFM}
    \xi(R)= \int_0^{\infty} P(k) \frac{\sin kR}{kR}\frac{k^2\mathrm{d}k}{2\pi^2} .
\end{equation}
For a statistical measurement of 2PCF in an N-point catalogue, a variety of edge-corrected estimators are proposed, among which the widely used estimator is given by \cite{LS1993} (hereafter LS). The symbolic expression for the LS-estimator is given by 
\begin{equation}\label{eq:LS-estimator}
\hat{\xi}_{\text{LS}} = \frac{DD-2DR+RR}{RR} ,
\end{equation}
where $DD$, $DR$ and $RR$ denotes for pair- or cross-pair-counting in the data $D$ and random sample $R$. We first consider the pair counting $DD$ in the data as a demonstration. $DD$ can be given by pair-counting in spherical shells with a separation distance $R$ and a width $\Delta R$. Let $n$ denote the number density, $DD $ can be given by the summation spreading over all the sampling particles
\begin{equation}\label{eq:pair-counting DD}
   DD(R,\Delta R) = \sum_{i}n({\bf x}_i)n_{R,\Delta R}({\bf x}_i) ,
\end{equation}
where $V_{R,\Delta R}$ is the volume of a spherical shell of finite width of $\Delta R$ at a radial distance $R$, $V_{R,\Delta R}= V_{R+\Delta R}-V_R$, in which $V_R= 4\pi R^3/3$ denotes the volume of a sphere of radius $R$, and $n_{R,\Delta R}({\bf x}_i)$ is the mean density within  the spherical shell $V_{R,\Delta R}$,  
\begin{equation}
   n_{R,\Delta R}({\bf x}_i) = \langle n({\bf x}_i+{\bf R})\rangle_{V_{R,\Delta R}} = \frac{1}{V_{R,\Delta R}}\int_{V_{R, \Delta R}} n({\bf x}_i+{\bf R}) d^3{\bf R}, 
\end{equation}
It is noted that we omit a common volume factor $V_{R,\Delta R}$ in the pair counts Eq.~\eqref{eq:pair-counting DD}, and thus $DD$ means pair-counting bi-density. In the following, we will keep this bi-density convention for pair counts. 

By introducing the top-hat window function of a spherical shell $W_{R, \Delta R}(\mathbf{x})$, we arrive at an equivalent representation by convolving the window function with the original density field, 
\begin{equation}\label{eq:den-Wshell}
    n_{R,\Delta R}({\bf x}) = W_{R, \Delta R}(\mathbf{x}) \circ n({\bf x}) = \int W_{R, \Delta R}(\mathbf{x-x}')n({\bf x}') d^3{\bf x}'.
\end{equation}
Obviously, the top-hat spherical shell in the radial range of $\{R, R+\Delta R\}$ can be written by the subtraction of two spherical top-hat windows, 
\begin{equation}\label{eq:window_shell_DeltaR}
W_{R, \Delta R}(\mathbf{x})=\frac{1}{V_{R, \Delta R}}(\theta(r-R-\Delta R)-\theta(r-R)) .
\end{equation}
Fourier-transforming the above equation yields the corresponding form of window function in the wavenumber space. Obviously, the window function of spherical shells with finite thickness in both real and wavenumber space can be written in a unified form by using its relation with the spherical top-hat, 
\begin{equation}\label{eq:window_shell_DeltaR_one}
W_{R, \Delta R}(\cdot)=\frac{1}{V_{R, \Delta R}}\left[V_{R+\Delta R} W_{\text {sphere }}(\cdot, R+\Delta R)-V_R W_{\text {sphere }}(\cdot, R)\right] ,
\end{equation}
where $\cdot$ denotes either $r$ in real space or $k$ in wavenumber space. 

Let $n({\bf x}) = \bar{n}(1+\delta({\bf x}))$, where $\bar{n}$ is a mean number density. According to the conventional pair-counting scheme described by Eq.~\eqref{eq:den-Wshell}, 2PCF under the working definition can be expressed by 
\begin{equation}\label{eq:2pcf-pow_shell}
    \xi_{\Delta R}(R) = \langle\delta({\bf x}), W_{R, \Delta R}(\mathbf{x}) \circ \delta ({\bf x})  \rangle.
\end{equation}
The Fourier relation between 2PCF and power spectrum Eq.~\eqref{eq:2PCFM} will be thereby altered to the following form
\begin{equation}\label{eq:2PCF-Pk-Wshell}
   \xi_{\Delta R}(R) = \int P(k)\hat{W}_{R, \Delta R}(k)\frac{k^2dk}{2\pi^2} 
\end{equation}

Under the thin spherical shell condition $\Delta_R = \Delta R/R \ll 1$ to the linear order, the window function Eq.~\eqref{eq:window_shell_DeltaR_one} in the k-space approximates to
\begin{equation}\label{eq:window_sph_expansion}
  \hat{W}_{R, \Delta R}(k) = (1-\eta_R)\frac{\sin kR}{kR} + \eta_R \cos kR, \quad \eta_R = \frac{1}{2} \frac{\Delta_R}{1+\Delta_R} ,
\end{equation}
where $\eta_R$ is a small weight factor. In the limit $k\rightarrow 0$, $\hat{W}_{R, \Delta R}(k) \rightarrow 1$ as required by the normalisation condition $\int W({\bf r})d^3{\bf r} =1$. 

Moreover, in the infinite thin shell limit $\Delta_R \rightarrow 0$,
Eq.~\eqref{eq:window_sph_expansion} reduces to 
\begin{equation}
    \tilde{W}_{R, \Delta R}({\bf k}) \rightarrow \tilde{W}_{\text{shell}}(k,R) =\frac{\sin(kR)}{kR} .
\end{equation}
which recovers the conventional expression Eq.~\eqref{eq:2PCFM}. 
Correspondingly, the window function in the real space becomes 
\begin{equation}\label{eq:sph-dirac}
W_{R, \Delta R}({\bf x}) \rightarrow {W}_{\text{shell}}(r,R) = \displaystyle{\frac{1}{4\pi r^2}}\delta_D(r-R) ,
\end{equation}
which has a straightforward geometric interpretation as the mean density within an infinitely thin spherical shell of radius $R$.

\begin{figure}
    \centering
    \includegraphics[width=6.0cm]{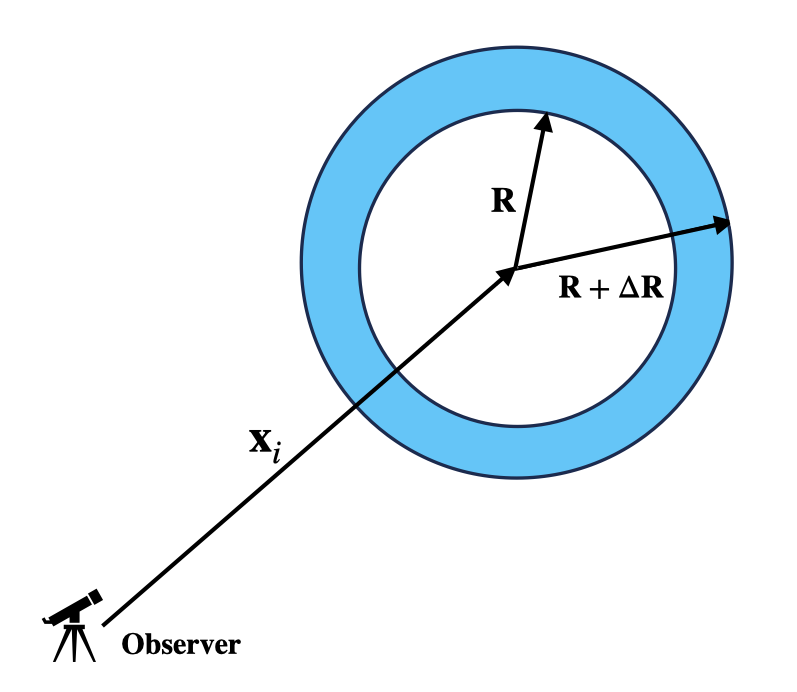}
%    \vspace{-0.2cm}
    \caption{Schematic plot of the spherical shell window given by Eq.(\ref{eq:window_shell_DeltaR}), which is a subtraction of two spherical tophats of radius $R$ and $R+\Delta R$.}
    \label{fig:sphshell-tophat}
\end{figure}

Similarly, according to the definition of the volume average 2PCF in spheres of radius $R$, we can simply get 
\begin{equation}\label{eq:2PCF-volume}
\begin{aligned}
\bar{\xi}(R) &= \langle\delta({\bf x}), W_{\text{sphere}}(\mathbf{x},R) \circ \delta ({\bf x})  \rangle \\
& = \int P(k)\hat{W}_{\text{sphere}}(k,R)\frac{k^2dk}{2\pi^2} ,
\end{aligned}
\end{equation}
where $\hat{W}_{\text{sphere}}(k,R)$ is the window function of a spherical tophat, 
\begin{equation}
\hat{W}_{\text {sphere }}(k, R)=\frac{3}{k^3 R^3}[\sin (k R)-k R \cos (k R)] .
\end{equation}

It is easy to find
\begin{equation}
    \frac{d}{dV_R}\Bigl[V_R W_{\text{sphere}}(\cdot,R)\Bigr] = W_{\text{shell}}(\cdot,R) ,
\end{equation}
and thus 
\begin{equation}\label{eq:tophat-shell-relation}
    \xi_{\Delta R}(R) = \displaystyle{\frac{1}{V_{R,\Delta R}}}\int_{V_{R, \Delta R}} \xi(R) dV_R.
\end{equation}
The above equation Eq.~\eqref{eq:tophat-shell-relation} has been used to account for the binning effect in the model analysis of the BAO signal from the SDSS DR7 data, e.g. \cite{Sanchez2008, Xu2012}.

\begin{figure*}
    \centering
    \includegraphics[width=\textwidth]{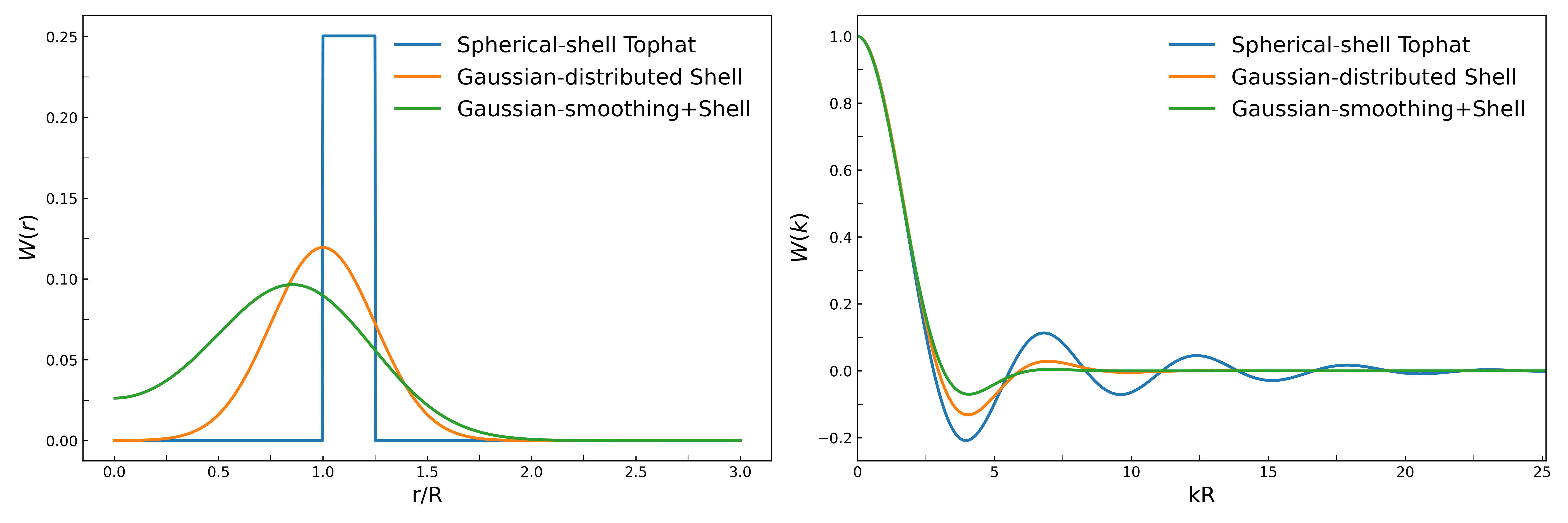}
%    \vspace{-0.5cm}
    \caption{The comparison of three typical window functions with finite width in real (left panel) and wavenumber space (right panel), including the spherical-shell tophat defined by Eq.~\eqref{eq:window_shell_DeltaR_one}, the Gaussian-distributed shell by Eqs.~(\ref{eq:GaussSmooth_R} -\ref{eq:GaussSmooth_k}), and the Gaussian-smoothing shell by Eqs.~(\ref{eq:Gauss+shell_k}-\ref{eq:Gauss+shell_R}). Varying the width of windows, we take $\Delta R/R = 0.25$ in the plot.}
    \label{fig1:kernel}
%    \label{fig:Gaussian kernel}
\end{figure*}

\subsection{Generalised 2PCF: from ex-situ to in-situ}
\label{sec:Generalised 2PCF}

Eq.~\eqref{eq:2pcf-pow_shell} gives an intuitive geometric generalisation of 2PCF to account for the binning in finite-thickness spherical shells. Mathematically, using a sharp-edged window function, a binned density field becomes equivalent to a filtered one. This equivalence allows the 2PCF to be extended beyond the traditional concept of pair counting, enabling more general and flexible forms of correlation function analysis. For an arbitrary window function, $W(\mathbf{x},\mathcal{P})$ characterized by a parameter set $\mathcal{P}$ of geometric configuration, the convoluted density field is thus,  
\begin{equation}
\delta_{\mathcal{P}}(\mathbf{x})=W(\mathbf{x}, \mathcal{P}) \circ \delta(\mathbf{x}) .
\end{equation}
Accordingly, the in-situ cross-correlation function of $\delta({\bf x})$ and $\delta_{\mathcal{P}}({\bf x})$ is, 
\begin{equation}
 \xi_{\mathcal{P}} =\langle\delta({\bf x})\delta_{\mathcal{P}}({\bf x})\rangle=\frac{1}{(2\pi)^3}\int P({\bf k})\hat{W}({\bf k},\mathcal{P})d^3{\bf k} .
\end{equation}
This generalisation provides flexibility in choosing a window function, and designing an appropriate one could be problem-dependent. Since we are trying to understand the clustering features, the window function selected should have a simple physical interpretation.

As a direct application, instead of the sharp Dirac function in the infinitely thin spherical shell window function Eq.~\eqref{eq:sph-dirac}, we make a natural extension by explicitly assigning a radial Gaussian profile with the filter radius $a$ at a distance of $R$ from a given particle,  
\begin{equation}
W_{\text {GS}}(r, R, a)=\frac{1}{\sqrt{32 \pi^3} a\left(R^2+a^2\right)}\left(e^{-(r-R)^2 / 2 a^2}+e^{-(r+R)^2 / 2 a^2}\right) ,
\label{eq:GaussSmooth_R}
\end{equation}
and its Fourier counterpart reads,
\begin{equation}
\hat{W}_{\text {GS}}(k, R, a)=\left[\frac{R^2}{R^2+a^2} \frac{\sin k R}{k R} + \frac{a^2}{R^2+a^2} \cos k R \right] e^{-\frac{1}{2} k^2 a^2} .
\label{eq:GaussSmooth_k}
\end{equation}
Clearly, except for the overall Gaussian smoothing, the finite radius of the Gaussian filter leads to an extra term of cosine, which is very similar to the tophat spherical shell window Eq.~\eqref{eq:window_sph_expansion}. In both cases, the appearance of the cosine component will capture extra phase information since cosine and sine are orthogonal to each other.    

An alternative extension could be given by applying a smoothing filter $W_{\mathcal{F}}$ in the original density field initially, 
\begin{equation}
    \delta_{\mathcal F} ({\bf x}) = W_{\mathcal{F}} \circ \delta({\bf x}) .
\end{equation}
In this case, the resulting 2PCF is thus 
\begin{equation}
\begin{aligned}
    \xi_{\mathcal{F}}(\mathcal{P})&=\langle \delta_{\mathcal{F}}({\bf x}), W_{\mathcal{P}}\circ\delta_{\mathcal{F}}({\bf x}) \rangle \\ 
    &=\langle W_{\mathcal{F}} \circ \delta({\bf x}), W_{\mathcal{F}}\circ W_{\mathcal{P}} \circ \delta({\bf x}) \rangle \\
    &=\langle \delta({\bf x}), W^{\dagger}_{\mathcal{F}}\circ W_{\mathcal{F}}\circ W_{\mathcal{P}} \circ \delta({\bf x}) \rangle \\
    &=\frac{1}{(2\pi)^3}\int P({\bf k}) |\hat{W}_{\mathcal{F}}({\bf k})|^2\hat{W}_{\mathcal{P}}({\bf k}){d^3{\bf k}} ,
\end{aligned}
\end{equation}
where the Dirac bra-ket notation in quantum mechanics has been used \citep[e.g.][]{dirac1949principles}, in which the window functions act as operators. This notation provides the advantage of facilitating transformations between various representations. 

A simple example can be illustrated by taking a Gaussian smoothing $\hat{W}_{\mathcal{F}} = \exp({-\frac{1}{2}k^2a^2})$ initially and then pair-counting by a spherical shell window $\hat{W}_{\mathcal{P}}=\hat{W}_{\text{shell}}$. Consequently, the combination effect will be given by $|\hat{W}_{\text{Gauss}}(k,a)|^2 \hat{W}_{\text{shell}}(k,R)$, that is, 
\begin{equation}
\begin{aligned}
    \hat{W}_{\text {Gauss+Shell }}(k, R, a) &=\left|\hat{W}_{\text {Gauss }}(k, a)\right|^2 \cdot \hat{W}_{\text {shell }}(k, R) \\
    &= \frac{\sin k R}{k R} e^{-k^2 a^2} .
\end{aligned} 
\label{eq:Gauss+shell_k}
\end{equation}
The corresponding window function in real space is found to be
\begin{equation}
    W_{\text {Gauss+Shell }}(r, R, a)=\frac{1}{8 \pi^{3 / 2} a R r}\left(e^{-(r-R)^2 / 4 a^2}-e^{-(r+R)^2 / 4 a^2}\right) .
    \label{eq:Gauss+shell_R}
\end{equation}
Fig.~\ref{fig1:kernel} displays the window functions of spherical-shell tophat, Gaussian spherical shell, and Gaussian-smoothing spherical shell. Obviously, the spherical-shell tophat has sharp, compact support in real space but is more extended in wavenumber space, while the Gaussian-distributed or Gaussian-smoothing shell has an extended distribution in real space and decays to zero more quickly beyond the sizes of Gaussian windows than the spherical-shell tophat in wavenumber space. It is noted that the Gaussian smoothing has been utilized to modify the BAO portion of the linear power spectrum only, accounting for the degradation of the BAO peak due to nonlinear effects and redshift distortions (e.g. \citealt{eisenstein2005detection, Tegmark2006, Crocce2006, eisenstein2007robustness, Crocce2008, matsubara2008resumming}). As indicated in Eq.~\eqref{eq:Gauss+shell_R}, this corresponds to attributing a Gaussian shape to the density wiggles, modulated by an amplitude decay of $1/r$ form in the BAO.

It is noted that we have only considered the 2PCF in real space, assuming spherical symmetry, which also applies to the introduced binning window function. However, the generalised 2PCF allows for window functions of arbitrary forms, tailored to match the observed symmetry of the matter distribution. For example, the spherical symmetry is broken in the redshift space and the 2PCF $\xi(\pi, \sigma)$ is typically parameterised by both the radial (redshift) distance $\pi$ and the transverse distances $\sigma$, which correspond to a ring-structured binning window function. Within the framework of the generalised 2PCF, the binning window function with an axially symmetric shape can be optimally chosen to improve redshift distortion measurements (Li et al. in preparation).

%%%%%%%%%%%%%%%%%%%%%%%%%%%%%%%%%%%%%%%%%%%%%%%%%%%%%%%%%%%%%%%%%%%%%%%%

\section{Fast Algorithm for 2PCF}

\subsection {The {\tt MRACS} algorithm for CIC}
\label{sec:MRACS}

In the preceding subsection, we made an algebraic extension of pair-counting from arithmetic pair-counting with top-hat windows to general continuous functions and further introduced the concept of generalised 2PCF. Here, we describe briefly a numerical scheme for performing a fast computation based on the {\tt MRCAS} algorithm \citep{Feng2007} implemented in the {\it Hermes} toolkit (Feng et al., in preparation). 

For a spatial point process, our {\tt MRACS} scheme is to convert the correlation between two points at a separation distance, i.e. $\langle n({\bf x})n({\bf x}+{\bf R})\rangle$ into the cross-correlation of two fields at the same point, $\langle n({\bf x})n_{\bf R}({\bf x})\rangle$, where
\begin{equation}
n_{\bf R}({\bf x})=W_{\bf R}({\bf x})\circ n({\bf x}) = \int W_{\bf R}({\bf x-x'})n({\bf x}')d^3{\bf x}' .
\end{equation}
Furthermore, a spatial point process can be modelled simply by 
\begin{equation}\label{eq:density_pointproc}
n({\bf x}) = \sum_i w_i \delta^3_D({\bf x}-{\bf x}_i), 
\end{equation}
where $w_i$ is an individual weight assigned to each particle, either a discrete or (truncated) continuous variable. For a galaxy sample, the weight could be a selection function (e.g. \citealt{Singh2021, Karim2023}), or a sort of mark related to the intrinsic properties of galaxies and their environments et al. as introduced in various marked clustering statistics (e.g. \citealt{Sheth2004, Sheth2005, Skibba2006, White2009, Simpson2011, Simpson2013, Skibba2013, White2016, Pujol2017, Aguayo2018, Armijo2018, Neyrinck2018, Valogiannis2018, Satpathy2019, Philcox2020, Alam2021_DESI, Massara2021, Xiao2022}). In general, the number density could be generalised to any physical fields associated with the spatial process, e.g. velocity or momentum field.

The continuous density field can be constructed by projecting Eq.~\eqref{eq:density_pointproc} onto a multiresolution space spanned by a set of complete and orthogonal basis functions $\phi_{j,{\bf l}}$ at a given resolution $j$, explicitly, 
\begin{equation}\label{eq:n-decomp}
  n^j({\bf x}) = \sum_{\bf l}\epsilon_{j,{\bf l}}\phi_{j,{\bf l}}({\bf x}), 
\end{equation}
where the scaling function coefficients (SFCs) $\epsilon_{j,{\bf l}}$ is given by
\begin{equation}\label{eq:scaling-coff}
   \epsilon_{j,{\bf l}}= \int n({\bf x})\phi_{j,{\bf l}}({\bf x})d^3{\bf x} = \sum_i w_i\phi_{j,{\bf l}}({\bf x}_i).
\end{equation}
Inserting Eq.(\ref{eq:scaling-coff}) into Eq.(\ref{eq:n-decomp}) yields
\begin{equation}
\begin{aligned}\label{eq:n-expansion}
    n^j({\bf x}) &=& \sum_{\bf l}\Bigl[\sum_{i}w_i\phi_{j,{\bf l}}({\bf x}_i)\Bigr]\phi_{j,{\bf l}}({\bf x}) \\
    &=& \sum_i w_i \Bigl[\sum_{\bf l} \phi_{j,{\bf l}}({\bf x}_i)\phi_{j,{\bf l}}({\bf x})\Bigr] .
    \end{aligned}
\end{equation}
Under the completeness condition of a set of basis functions in the infinite limit of $j\rightarrow\infty$, the in-bracket term goes to
\begin{equation}
\sum_{\mathbf{l}} \phi_{j 1}(\mathbf{x}) \phi_{j 1}\left(\mathbf{x}^{\prime}\right)=\Delta_j\left(\mathbf{x}, \mathbf{x}^{\prime}\right) \rightarrow \delta_D\left(\mathbf{x}-\mathbf{x}^{\prime}\right).
\end{equation}
Eq.(\ref{eq:n-expansion}) implies that we have used essentially a set of basis functions to approximate the singular Dirac function $\delta_D$ asymptotically with increasing $j$. Mathematically, the basis functions are required to have both a compact support and a high convergence rate in the sense of completeness. Reconstruction of the continuous density field makes numerical computation possible. The compactly supported basis functions widely used in cosmic statistics include families of B-spline (e.g. \citealt{Feng2007,Yang2010}), Daubechies scaling functions (e.g. \citealt{Fang1998Book,Feng2000,yang2001a,yang2001b,yang2002,Yang2003,Cui2008}), etc.    

We notice that, given that the scaling functions $\{\phi_{j,{\bf l}}\}$ are generated via dilation by $2^j$ and translation by ${\bf l}$ of the father function $\phi({\bf x})$, which typically has compact support. Therefore, summing up nonzero contributions in Eq.~\eqref{eq:scaling-coff} comes solely from nearby particles within the support domain centred on the position ${\bf l}$.

Generally, a window function could have a bi-linear decomposition in terms of the basis function, 
\begin{equation}
W({\bf x},{\bf y})=\sum_{\bf l,m}\varpi^j_{{\bf l},{\bf m}}\phi_{j,{\bf l}}({\bf x})\phi_{j,{\bf m}}({\bf y}).
\end{equation}
Accordingly, the CIC field $n_{\bf R}({\bf x})$ can be written by 
\begin{equation}\label{eq:nr-decomp}
  n_{\bf R}({\bf x}) = \sum_{\bf l}\epsilon_{j,{\bf l}}^{\bf R}\phi_{j,{\bf l}}({\bf x}),
\end{equation}
where 
\begin{equation}\label{eq:convolutedSFC}
\epsilon_{j,{\bf l}}^{\bf R} = \sum_{\bf m} \varpi^j_{{\bf l},{\bf m}}\epsilon_{j,{\bf m}}.
\end{equation}
For a homogeneous kernel, $\varpi^j_{{\bf l},{\bf m}} = \varpi^j_{{\bf l}-{\bf m}}$ is the Toeplitz matrix; conventionally, the above matrix multiplication can be accomplished using the FFT technique. A CIC numerical experiment for a given spatial configuration requires a pair of FFT and inverse FFT computations, and its complexity scales to $N_g\log N_g$ ($N_g$ is the number of FFT grids). Obviously, once the convoluted SFCs Eq.(\ref{eq:convolutedSFC}) are obtained, the complexity of reading out the CIC field (\ref{eq:nr-decomp}) at any point is only $\mathscr{O}(D_{\text{supp}}^3)$, where $D_{\text{supp}}$ is the support size of the basis functions. Typically, it is almost an algorithm at the $\mathscr{O}(1)$ level. 

\subsection{Fast algorithm for 2PCF: an in-situ approach}

Under the ergodic assumption, the pair counts $DD$ can be equivalently expressed by a volume average, that is, 
\begin{equation}\label{eq:DD-ensemble}
  DD = \langle n({\bf x})n_{\bf R}({\bf x})\rangle = \frac{1}{V}\int_V n({\bf x})n_{\bf R}({\bf x})d^3{\bf x}. 
\end{equation}
where $n_{\bf R} = W \circ n$ is the binned density field made by convolving the original density field with a binning window function $W$. Following the conventional definition of 2PCF in the real space, discrete pair counts can only be done within a finite-width spherical shell centred around a given point, the window function $W$ is parameterised by both the radius and width of the shell, as given in Eq.~\eqref{eq:window_shell_DeltaR_one}.   

For the density field $n({\bf x})$ modelled by a point process Eq.~\eqref{eq:density_pointproc}, the above spatial average Eq.~\eqref{eq:DD-ensemble} can be expressed in a summation over all the sampling particles by using direct integration,
\begin{equation}
    DD = \frac{1}{V}\sum_i n_{\bf R}({\bf x}_i) 
\end{equation}
which recovers the usual meaning of pair counts. Furthermore, when
substituting Eq.~\eqref{eq:n-decomp} and Eq.~\eqref{eq:nr-decomp} into the above equation Eq.~\eqref{eq:DD-ensemble}, we arrive at
\begin{equation}\label{eq:DD-fast}
DD=\sum_{\mathbf{l,m}} {\epsilon}_{j,{\bf l}} {\epsilon}_{j,{\bf m}}^{\bf R} \int \phi_{j,{\bf l}}(\mathbf{x}) \phi_{j,{\bf m}}(\mathbf{x}) d^3 \mathbf{x}=\sum_{\mathbf{l}}{\epsilon}_{j,{\bf l}}{\epsilon}^{\bf R}_{j,{\bf l}},
\end{equation}
where the orthogonality of basis functions has been used. According to Eq.~\eqref{eq:DD-fast}, the pair-counting is simplified by a scalar product of the scaling function coefficients of two fields. Moreover, if we are only interested in measuring the 2PCF, the above sum can be performed in the wavenumber space, as ensured by Parseval's theorem, which only requires performing the FFT once on the scaling coefficients $\epsilon_{j{\bf l}}$, regardless of the number of radial bins we applied. On other hand, since FFT calculations are computationally inexpensive on most standard platforms, if we wish to store the filtered density fields for other purposes, we will need to perform an additional inverse FFT calculation for each radial bin of interest.      

It is instructive to see that binning can be effectively modelled by a window function, which equates the pair-counting process to CIC or filtering operation.  As discussed above, by choosing a window function $W = W_{R,\Delta R}$, one can replicate the procedure of pair-counting within finite-width spherical shells. However, the pair count given by Eq.~\eqref{eq:DD-fast} suggests that the binning window function can be generalised to any low-pass filter without limiting itself to discrete counts. In particular, Eq.~\eqref{eq:DD-fast} still works even in the limit of infinitesimal binning $\Delta R \rightarrow 0$, $W\rightarrow W_{\text{shell}}$, although the traditional counting method is no longer applicable. This insight is essential to understanding the computational feasibility of the generalised 2PCF proposed in Section \ref{sec:Generalised 2PCF}.     

Remarkably, an in situ perspective of 2PCF in the {\tt MRACS} scheme leads to a fast algorithm to estimate 2PCF in large data sets. By incorporating binning corrections into the theoretical model, this approach establishes a direct connection between theoretical predictions and practical 2PCF measurements based on the generalised binning scheme. 

In our statistical analysis, we continue to use the edge-corrected estimator to measure the 2PCF in simulation samples. Specifically, we employ the Landy-Szalay (LS) estimator given by Eq. (\ref{eq:LS-estimator}) \citep{LS1993}, which minimises the shot noise to the second order and can be extended to higher order correlation functions \citep{SzapudiSzalay1998}. In the abbreviated symbolic expression, the LS estimator can be expressed as
\begin{equation}
\hat\xi_{LS} = \frac{\langle (D-R)^2 \rangle } {\langle R^2 \rangle}
\end{equation}
For a given random sample R, we follow the same procedure as for the data sample to calculate its scaling coefficients (SFC) $\epsilon_{j{\bf l}}^R$. The difference $(D-R)$ is then evaluated by subtracting the SFCs of the random sample from those of the data, yielding $\Delta\epsilon_{j{\bf l}} = \epsilon_{j{\bf l}} - \epsilon_{j{\bf l}}^R$. Consequently, $\langle (D-R)^2 \rangle$ can be estimated using the same summation rule as in Eq. (\ref{eq:DD-fast}).

\section{Filtered 2PCF and Numerical Test} 

\begin{figure*}
    \centering
    \includegraphics[width=\textwidth]{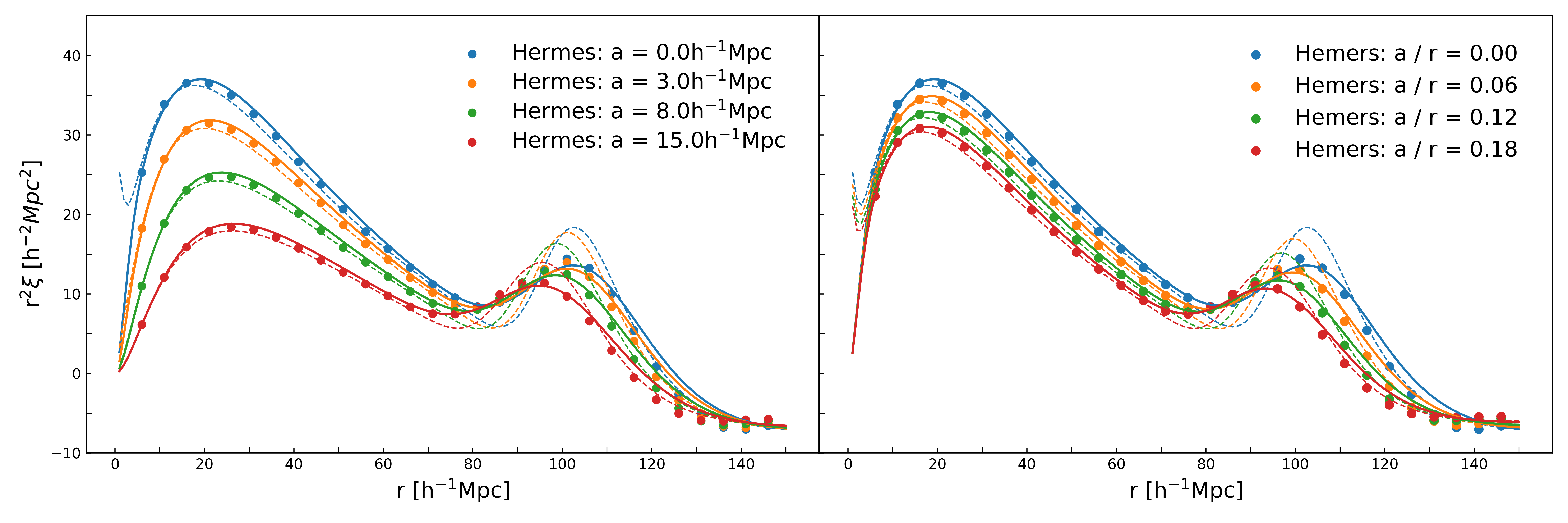}
    \vspace{-0.5cm}
    \caption{Plot of the 2PCF measured in the BigMDPL simulation, in which the binning is using the spherical shell tophat given by Eq.~\eqref{eq:window_shell_DeltaR} with fixed bin sizes (left) and varying binning linearly scaled with radial distance (right). In the former, bin sizes are taking (0, 3, 8, 15) $h^{-1}$Mpc, while in the latter, ratios of bin size to radius are set to 0.00, 0.06, 0.12, and 0.18, respectively. In comparison, we plot the corresponding theoretical predictions derived from the power spectra under the Zeldovich approximation (solid lines) and the nonlinear power spectra (thin dashed lines), which are output from the Nbodykit package.}
    \label{fig:eq3 comparison}
\end{figure*}

\subsection{Description of simulation samples}

In this study, we perform clustering analysis in two publicly available simulation suites: MultiDark simulations\footnote{https://www.multidark.org} and Quijote simulations\footnote{https://quijote-simulations.readthedocs.io/en/latest/}. Our numerical experiment serves a dual purpose. First, we aim to validate the generalised correlation functions defined by a set of filters and corresponding binning corrections using high-resolution simulations. Second, we investigate the impact of binning on statistical variances by analysing a series of simulation realisations.

The BigMultiDark Planck \citep{prada2012halo,riebe2013multidark} (BigMDPL) simulation evolves $3840^3$ dark matter particles in a box of side length of 2500$h^{-1}$ Mpc, which has a mass resolution of $2.359\times 10^{10} h^{-1}$ M\textsubscript{\(\odot\)}. This large volume allows the 2PCF to be measured up to 150$h^{-1}$ Mpc with high statistical precision. BigMDPL assumes a flat cold dark matter ($\Lambda$CDM) cosmological model, which is normalised by the Planck13 parameters: ($h$, $\Omega_m$, $\Omega_b$, $n_s$, $\sigma_8$) = (0.6777, 0.307115, 0.048206, 0.96, 0.8228). Alternatively, we used MultiDark Planck 2 (MDPL2) in the 3PCF analysis. The MDPL2 simulation has the same cosmological parameters and particle number as the BigMDPL simulation but a smaller box size of 1000$h^{-1}$ Mpc, reaching both high spatial and mass resolutions. To save storage, we extract two sub-samples with a sampling rate of $0.5\%$ from the original BigMDPL and MDPL2 snapshots at redshift $z=0$, both of which comprise $\sim 2.8\times 10^8$ dark matter particles, though the time complexity of the {\tt MRACS} algorithm is almost independent of the number of particles. 

Another set of simulations we analysed is from the Quijote project \citep{villaescusa2020quijote}, which releases a suite of more than 82,000 full N-body simulations designed to quantify the information content on cosmological observables. The Quijote simulation provides a large number of realisations with a box size of 1000$h^{-1}$ Mpc, containing $512^3$ dark matter particles and around $4\times 10^5$ halos per realisation. We used simulation samples in Planck18 cosmology specified by the following parameters: ($h$, $\Omega_m$, $\Omega_b$, $n_s$, $\sigma_8$) = (0.6711, 0.3175, 0.049, 0.9624, 0.834), in which the Latin hypercube sampling technique was used to generate the initial conditions. This large number of realisations allows for robust measurements of the cosmic variance of the 2PCF and 3PCF. Since this paper makes no attempt to assess the statistical precision in cosmological measurements, we select only 50 realisations for dark matter and the corresponding halo samples. 

\subsection{2PCF: Numerical test}
\label{sec:2PCFtest}

\begin{figure*}
    \centering
    \includegraphics[width=\textwidth]{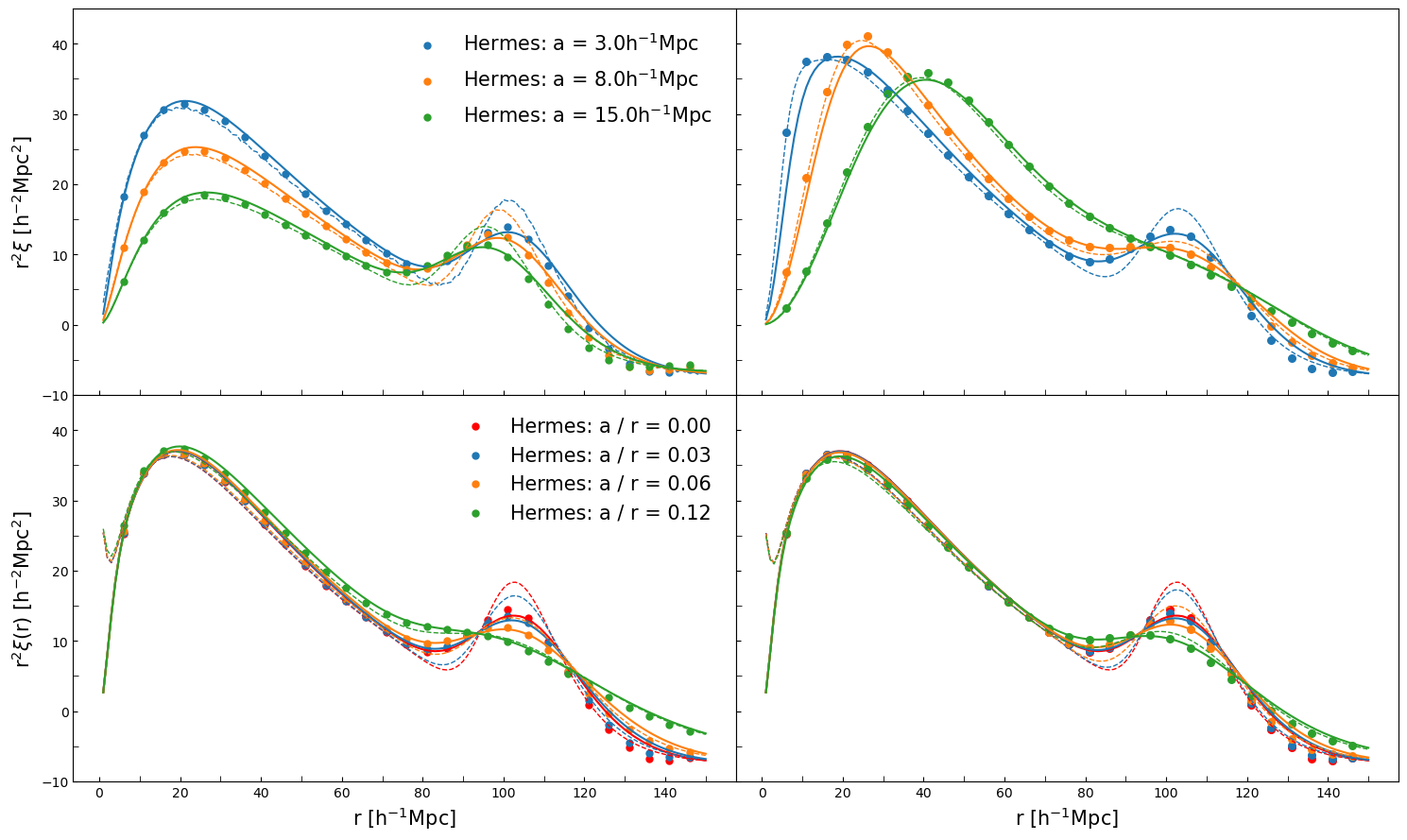}
    \vspace{-0.5cm}
    \caption{Same as Fig.~\ref{fig:eq3 comparison} but for the Gaussian-distributed shell Eq.~\eqref{eq:GaussSmooth_k} (left) and Gaussian-smoothing+shell Eq.~\eqref{eq:Gauss+shell_k} (right) with the fixed bin sizes (3, 8, 15) $h^{-1}$Mpc (upper), and varying bin sizes with the ratios $a/r$ of 0, 0.03, 0.06 and 0.12 (lower).}
    \label{fig:Gaussian kernel}
\end{figure*}

\begin{figure*}
    \centering
    \includegraphics[width=\textwidth]{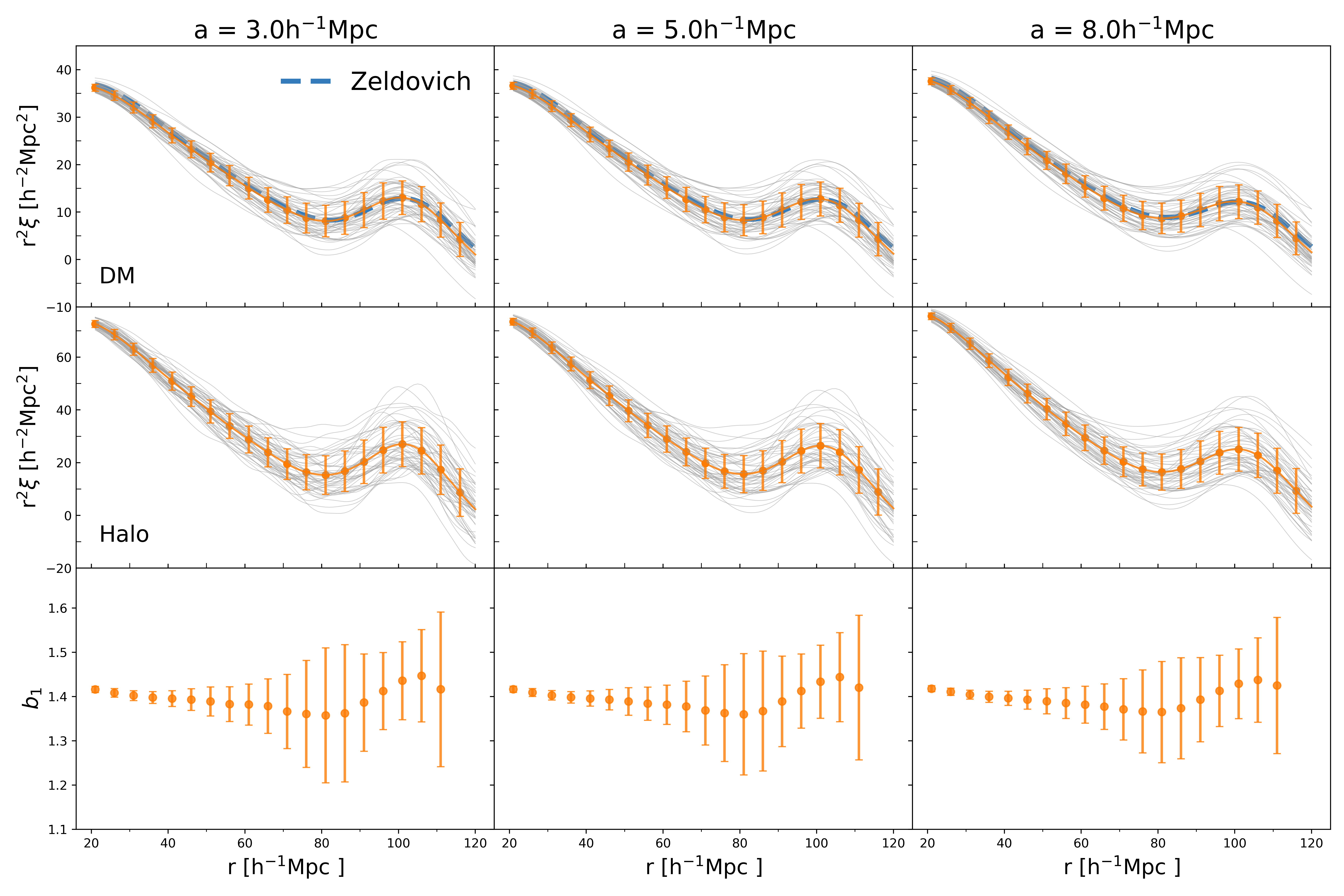}
    \vspace{-0.5cm}
    \caption{The 2PCF with the spherical tophat filter measured in the 50 realizations of the Quijote simulations. The binning is specified by the radius of the spherical tophat, which is set to 3 (left), 5 (middle), and 8 (right) $h^{-1}$Mpc in three columns, respectively. The auto-correlation functions of dark matter particles and halos are presented in the upper and middle rows, respectively, and the corresponding linear bias parameters obtained by $b_1 = \sqrt{\xi_{hh}/\xi_{mm}}$ with 1-$\sigma$ error bars are plotted in the lower row. In all the plots, the solid lines indicate the mean values over the 50 realizations, and the error bars on the data points represent 1-$\sigma$ dispersion. In the top panel for 2PCF of dark matter particles, the theoretical curves in the Zeldovich approximation are also plotted by dash lines for comparison with the simulations.}
    \label{fig:quijote 2pcf}
\end{figure*}

Throughout the 2PCF analysis performed on the simulation samples, we reconstruct the density fields using a decomposition in terms of Daubechies-4 scaling functions on a dilation scale of $J=10$ and $8$, for the BigMDPL and Quijote simulations, respectively. For the BigMDPL simulation, this choice of $J=10$ corresponds to a spatial resolution of $2500/2^{10}=2.44h^{-1}$Mpc, while for the Quijote simulations with the box size of $1 h^{-1}$ Gpc, the corresponding spatial resolution is $3.91h^{-1}$Mpc at $J=8$. 

Using the current pair counts based on spherical-shell tophat with finite widths, we first measure the 2PCF in the BigMDPL dark matter sample. We adopted two different binning schemes, fixed shell widths (left) and varying shell widths adaptive to radial distances. The results are displayed in Fig.~\ref{fig:eq3 comparison}. Larger bin sizes will significantly suppress the shot noise, leading to stronger smoothing on comparable scales. In the case of fixed widths, when the spherical shell width takes a relatively large value, e.g. $a \ge 8 \, h^{-1} \text{Mpc}$,  there will be over-smoothing on small scales, resulting in a marked reduction in the signal. For instance, $r^2 \xi$ around the first peak $\sim 20h^{-1} \text{Mpc}$ is reduced by approximately $30\%$ at a bin size of $a = 8 h^{-1} \text{Mpc}$ and by approximately $50\%$ at $a = 15 h^{-1} \text{Mpc}$. In contrast, when varying shell widths are employed, the binning adapts to the radial distance between two points, meaning that larger bin sizes are applied to the density field on larger scales. As a result, the overall shape of the 2PCF is not significantly distorted by the binning effect, although the 2PCF profiles do shift slightly towards smaller scales as the ratio $a/r$ increases. This adaptive approach mitigates over-smoothing, preserving the signal on small scales while still benefiting from reduced shot noise on larger scales.

For comparison with model predictions, we also plot the theoretical curves derived from the non-linear halofit model \citep{smith2003stable} and the analytical model based on the Zeldovich approximation, both of which are computed using the open source {\tt nbodykit} package \citep{Hand2018}. As anticipated, on scales around the BAO peak, the quasilinear model grounded in the Zeldovich approximation shows superior agreement with the simulations. In contrast, the nonlinear model provides a good fit on both small nonlinear scales and large scales beyond the BAO peak. However, it is noteworthy that the nonlinear fitting model begins to diverge from the simulation data near the first trough in the 2PCF. Despite this, it exhibits an excellent agreement gain on larger linear scales, where linear perturbation theory remains valid. This behaviour highlights the distinct regimes where each model excels: the Zeldovich approximation on quasilinear scales near the BAO peak and the nonlinear model on both smaller, nonlinear scales and larger, linear scales.

In Section 2, we introduced a generalised concept of the 2PCF, which extends beyond the traditional approach of discrete pair counting within spherical shells. To illustrate the generalised 2PCF, we present the results for two typical binning schemes applied to the BigMDPL simulation, as shown in Fig.~\ref{fig:Gaussian kernel}. In the upper left panel, we display the 2PCF measured using a Gaussian-distributed shell. The Gaussian shell, as a straightforward extension of the Dirac delta distribution, exhibits behaviours similar to those of the Dirac function, with a peak centred around the mean separation. However, when we examine the Gaussian-smoothing density field in the upper right panel of Fig.~\ref{fig:Gaussian kernel}, we observe a significant flattening of the 2PCF as the filtering scale increases, indicating the smoothing effect of the Gaussian filter.

Moreover, we provide results using adaptive binning with linear factors of 0.03, 0.06, and 0.12, alongside a null binning result for comparison. Unlike the spherical-shell tophat results shown in Fig. \ref{fig:eq3 comparison}, where binning has a noticeable impact on the 2PCF’s overall shape, the adaptive binning in the Gaussian-smoothing case shows less influence on the overall shape, with deviations primarily occurring around the BAO peak. This suggests that the smoothing and binning effects are more nuanced and dependent on the specific filter applied.

As discussed earlier, binning introduces a trade-off in the clustering analysis of spatial point processes. Smaller bin sizes reduce bias but amplify fluctuations in the signal, while larger bin sizes smooth out fluctuations but introduce bias. However, errors due to cosmic variance remain unaffected by the choice of binning. Figure \ref{fig:quijote 2pcf} illustrates the impact of binning on the estimations of the autocorrelation functions for dark matter particles and halos, as well as the cross-correlation function between them, based on 50 realisations from the Quijote simulation suite.

For the halo samples, with number densities around $\sim 4.0\times 10^{-4}(h^{-1} \text{Mpc})^{-3}$ - approximately $0.3\%$ of the density of dark matter - the density field was smoothed using a spherical tophat filter rather than a Gaussian filter, resulting in an effective window function of $\hat{W}_{\text{sphere}}\hat{W}_{\text{shell}}$. As shown in Figure \ref{fig:quijote 2pcf}, increasing the smoothing radius, that is, increasing the binning, will average more neighbour data points and suppress systematic errors, especially the fraction of shot noise on small scales due to sparse sampling, producing smoother 2PCF shapes in halo catalogues \citep{Cohn2006, Smith2009}. Additionally, the 1$\sigma$ variances remain largely unchanged with increasing filter radius on large scales, indicating that the statistical errors due to cosmic variance across different realisations are consistent, regardless of the smoothing applied.   

In the bottom row of Fig.~\ref{fig:quijote 2pcf}, we also present the scale dependence of the linear bias parameter $b_1$, derived from the preceding two rows using the relation $b_1 = \sqrt{\xi_{hh}/\xi_{mm}}$. In particular, across the scales ranging from $20 h^{-1}\text{Mpc}$ to $80 h^{-1}\text{Mpc}$ - specifically away from the BAO peak — $b_1$ shows only a slight decline with increasing scale. For instance, on the filter scale of $a = 3 h^{-1}\text{Mpc}$, $b_1$ varies from $1.416$ to $1.361$, corresponding to a fractional change of less than $3.9\%$. Beyond this range, a reverse trend appears, leading to a noticeable increase in $b_1$, which reaches a maximum value of $b_1=1.45$ near the BAO peak. Moreover, as shown in Fig.~\ref{fig:quijote 2pcf}, $b_1$ displays a weak dependence on the filtered scales.   

\section{Extending to Three-Point Correlation Function}

\subsection{Quantifying the binning effect in 3PCF}
\label{sec:3PCF-theory}

The 2PCF can be straightforwardly generalised to any higher order $N>2$, which is defined by the joint probability of finding objects in $N$ infinitesimal volumes. For $N=3$, the 3PCF can be yielded through 
\begin{equation}
\begin{aligned}
d P_{123}= & \bar{n}\left(\mathbf{r}_1\right) \bar{n}\left(\mathbf{r}_2\right) \bar{n}\left(\mathbf{r}_3\right)\left[1+\xi\left(r_{12}\right)+\xi\left(r_{23}\right)\right. \\
& +\xi\left(r_{31}\right)+\zeta\left(r_{12}, r_{23}, r_{31}\right) ] d^3 \mathbf{r}_1 d^3 \mathbf{r}_2 d^3 \mathbf{r}_3 .
\end{aligned}
\end{equation}

The 3PCF and the bispectrum are related through a Fourier transform., 
\begin{equation}\label{eq:3pcf}
\zeta({\bf r}_1, {\bf r}_2,{\bf r}_3) = \frac{1}{(2\pi)^6}\int \prod_i d^3{\bf k}_iB({\bf k}_1,{\bf k}_2,{\bf k}_3)\delta_D^3(\sum_i{\bf k}_i) e^{i\sum_i{\bf k}_i\cdot{\bf r}_i} .
\end{equation}
The bispectrum can be parameterized by two lengths $k_1$, $k_2$, and the angle between them and could be decomposed in terms of spherical harmonics
\citep{Szapudi2004},
\begin{equation}\label{eq:bipow-sph}
B(k_1,k_2,\mu) = \sum_l \frac{2l+1}{4\pi} B_l(k_1,k_2)P_l(\hat{\bf k}_1\cdot\hat{\bf k}_2) .
\end{equation}

In practical measurements of the 3PCF, the intensive computation arises from the triplet counting for a given triangle configuration. Recall that the {\tt MRACS} algorithm facilitates rapid readout of the CIC at any spatial location. Accordingly, we adopt the following binning scheme: for each vertex of a given shape triangle, we apply an identical spherical tophat window with a given filtered radius $R$ to produce the binned density field, $n_R({\bf x}_i) = W_{\text{sphere}}({\bf x}, R) \circ n({\bf x}_i), ~ i=1,2,3$. This approach modifies the bispectrum in Eq.~\eqref{eq:bipow-sph} by incorporating a multiplication of the three window functions in wavenumber space.
\begin{equation}
    \hat{W}_{\text{sphere}}(k_1,R) \hat{W}_{\text{sphere}}(k_2,R) \hat{W}_{\text{sphere}}(|{\bf k}_1+{\bf k}_2|,R).
\end{equation}
Clearly, the angle dependence is from the term $ W_{\text{sphere}}(|{\bf k}_1+{\bf k}_2|,R)$ with $|{\bf k}_1+{\bf k}_2| = [k_1^2 + k_2^2 + 2k_1k_2(\hat{\bf k}_1\cdot\hat{\bf k}_2)]^{1/2}$. To obtain an expansion proceeding in a series of the Legendre polynomials, we go back to its definition of the Fourier transformation. 
\begin{equation}
    \hat{W}_{\text {sphere }}\left(\left|\boldsymbol{k}_1+\boldsymbol{k}_2\right|, R\right)=\frac{3}{4\pi R^3}\int \mathrm{d}^3 \boldsymbol{r} \theta(R-r) e^{i\left(\boldsymbol{k}_1+\boldsymbol{k}_2\right) \cdot \boldsymbol{r}} .
\end{equation}
On using the expansion 
\begin{equation}
    e^{i \mathbf{k} \cdot \mathbf{r}}=4 \pi \sum_{l=0}^{\infty} \sum_{m=-l}^l i^l j_l(k r) Y_l^m\left(\Omega_k\right)^* Y_l^m\left(\Omega_r\right),
\end{equation}
and the spherical harmonic addition theorem, 
\begin{equation}
    P_l(\hat{\bf n}_1\cdot\hat{\bf n}_2)=\frac{4 \pi}{2 l+1} \sum_m Y_l^m\left(\hat{\bf n}_1\right)^* Y_l^m\left(\hat{\bf n}_2\right),
\end{equation}
a straightforward calculation yields
\begin{equation}\label{eq:tophat_legendre}
    \hat{W}_{\text{sphere}}(|{\bf k}_1+{\bf k}_2|,R) =\sum_m (-1)^m (2m+1) G_m(k_1,k_2,R)P_m(\hat{\bf k}_1\cdot\hat{\bf k}_2)
\end{equation}
with 
\begin{equation}
\begin{aligned}
&G_m\left(k_1, k_2, R\right) =\frac{3}{4\pi R^3}\int_0^R j_m\left(k_1 r\right) j_m\left(k_2 r\right) 4 \pi r^2 \mathrm{~d} r \\
& =\frac{3\pi}{2R^2} \frac{k_2 J_{m-\frac{1}{2}}\left(k_2 R\right) J_{m+\frac{1}{2}}\left(k_1 R\right)-k_1 J_{m-\frac{1}{2}}\left(k_1 R\right) J_{m+\frac{1}{2}}\left(k_2 R\right)}{\left(k_1^2-k_2^2\right)\left(k_1 k_2\right)^{\frac{1}{2}}}. 
\end{aligned}
\end{equation}

\begin{figure}
    \centering
    \includegraphics[width=8.0cm]{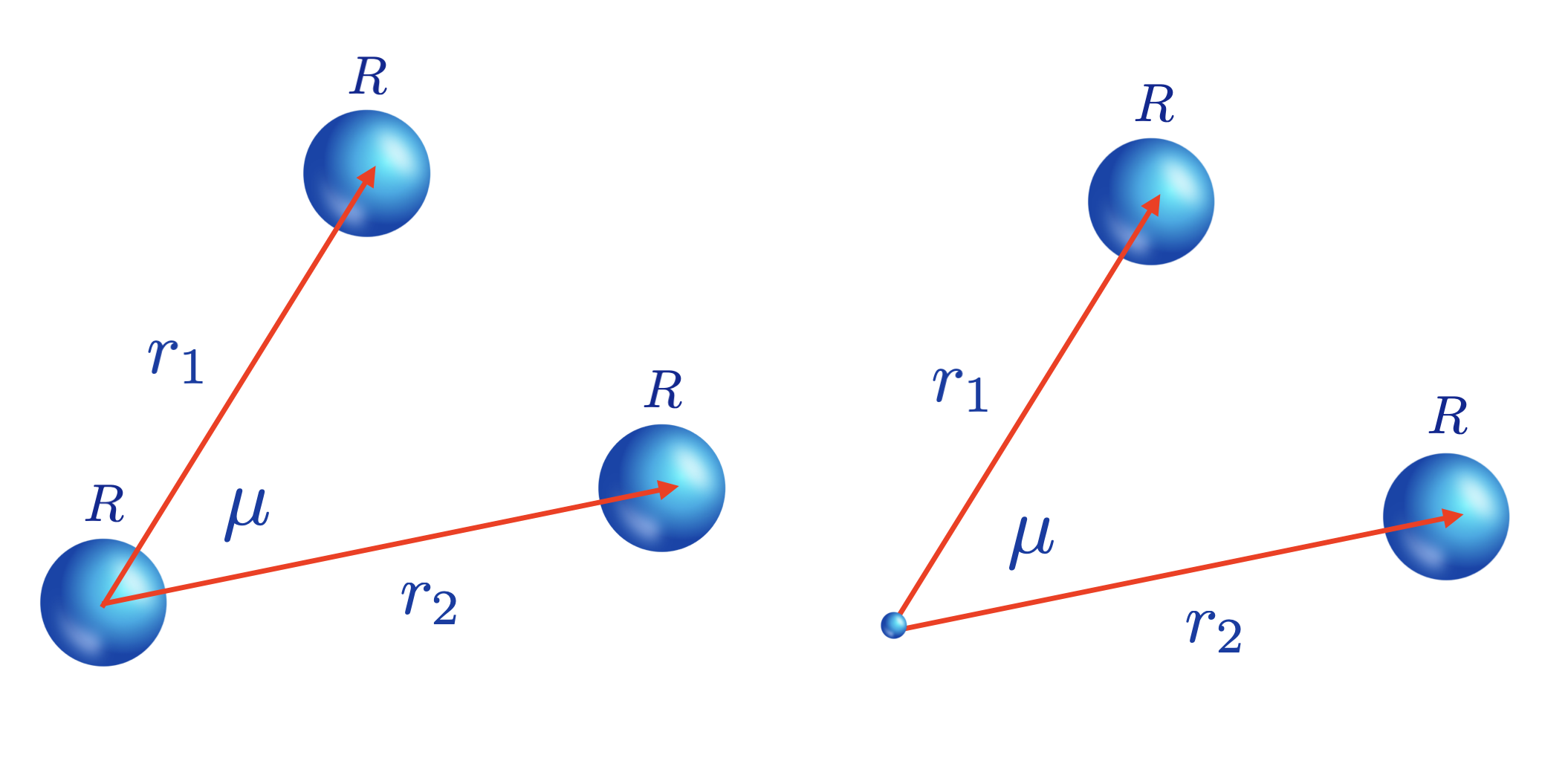}
    \vspace{-0.3cm}
    \caption{Schematic plot for measuring the 3PCF in data-catalog utilizing the triplet counting in 3-spheres. The left panel: in the meaning of Monte-Carlo integration for a spatial average over whole volume, we make CIC in three spheres at three vertices of a given triangle configuration, which are randomly placed in the whole space. The right panel: the first vertex of triangles goes through each object in the catalogue, and the CIC is only made in 2-spheres in the other two vertices.}
    \label{fig:3pcf-config}
\end{figure}

Combining Eqs.~(\ref{eq:3pcf}),(\ref{eq:bipow-sph}) and (\ref{eq:tophat_legendre}), and using the relation
\begin{equation}
    P_l P_m=\sum_{n=|l-m|}^{l+m}\left(\begin{array}{lll}
l & m & n \\
0 & 0 & 0
\end{array}\right)^2{(2n+1) P_n},
\end{equation}
we obtain the filtered 3PCF in the following form
\begin{equation}\label{eq:3pcf-tophat}
\zeta(r_1,r_2,\hat{\bf r}_1\cdot \hat{\bf r}_2)  = \sum_{n}\zeta_n(r_1,r_2) P_n(\hat{\bf r}_1\cdot \hat{\bf r}_2) 
\end{equation}
in which
\begin{equation}\label{eq:3pcf-tophat-coff}
\begin{aligned}
&\zeta_n(r_1,r_2)  = \sum_{lm}C_{lmn}\int\frac{k_1^2dk_1}{2\pi^2}\frac{k_2^2dk_2}{2\pi^2}j_n(k_1r_1)j_n(k_2r_2) \\ &\hat{W}(k_1,R)\hat{W}(k_2,R)G_m(k_1,k_2,R)B_l(k_1,k_2),
\end{aligned}
\end{equation}
with
\begin{equation}
C_{lmn} = (-1)^{m+n}\frac{1}{4\pi} (2l+1)(2m+1)(2n+1)\left(\begin{array}{lll}
l & m & n \\
0 & 0 & 0
\end{array}\right)^2 .
\end{equation}

Eq.~\eqref{eq:3pcf-tophat} can also be extended for the Gaussian window function by modifying the function $G_m$ with
\begin{equation}\label{eq:Gm-Gauss}
    G_m(k_1,k_2,R)=\sqrt{\frac{\pi}{2k_1k_2R^2}}\exp\Bigl\{{-\frac{(k_1^2+k_2^2)R^2}{2}}\Bigr\}{\text{I}}_{m+\frac{1}{2}}(k_1k_2R^2), 
\end{equation}
where ${\text{I}}_{m+\frac{1}{2}}(\cdot)$ are modified Bessel functions.
In the derivation of Eq.~\eqref{eq:Gm-Gauss}, the following expansion \citep{Bernardeau2002}
\begin{equation}
    \mathrm{e}^{-p q \cos \theta}=\sum_{m=0}^{\infty}(-1)^m(2 m+1) \sqrt{\frac{\pi}{2 p q}} I_{m+1 / 2}(p q) P_m(\cos \theta),
\end{equation}
has been applied for decomposing the angular part in the Gaussian window function $\hat{W}_{\text{Gauss}}(|{\bf k}_1+{\bf k}_2|,R)$. 

In the theoretical expressions for the filtered 3PCF, Eq.~\eqref{eq:3pcf-tophat} and Eq.~\eqref{eq:3pcf-tophat-coff}, we have assumed identical window functions, either spherical tophat or Gaussian. However, this assumption can be relaxed, as suggested by the above derivation. It is feasible to assign different spherical window functions — varying in shape or filtered radius — to each vertex of the triangles. Adjusting the current formulae to accommodate this variation is straightforward and allows for greater flexibility in the analysis.

\subsection{Triple sphere binning in 3PCF measurements}
\label{sec:3PCF-binning}

\begin{figure*}
    \centering
    \includegraphics[width=\textwidth]{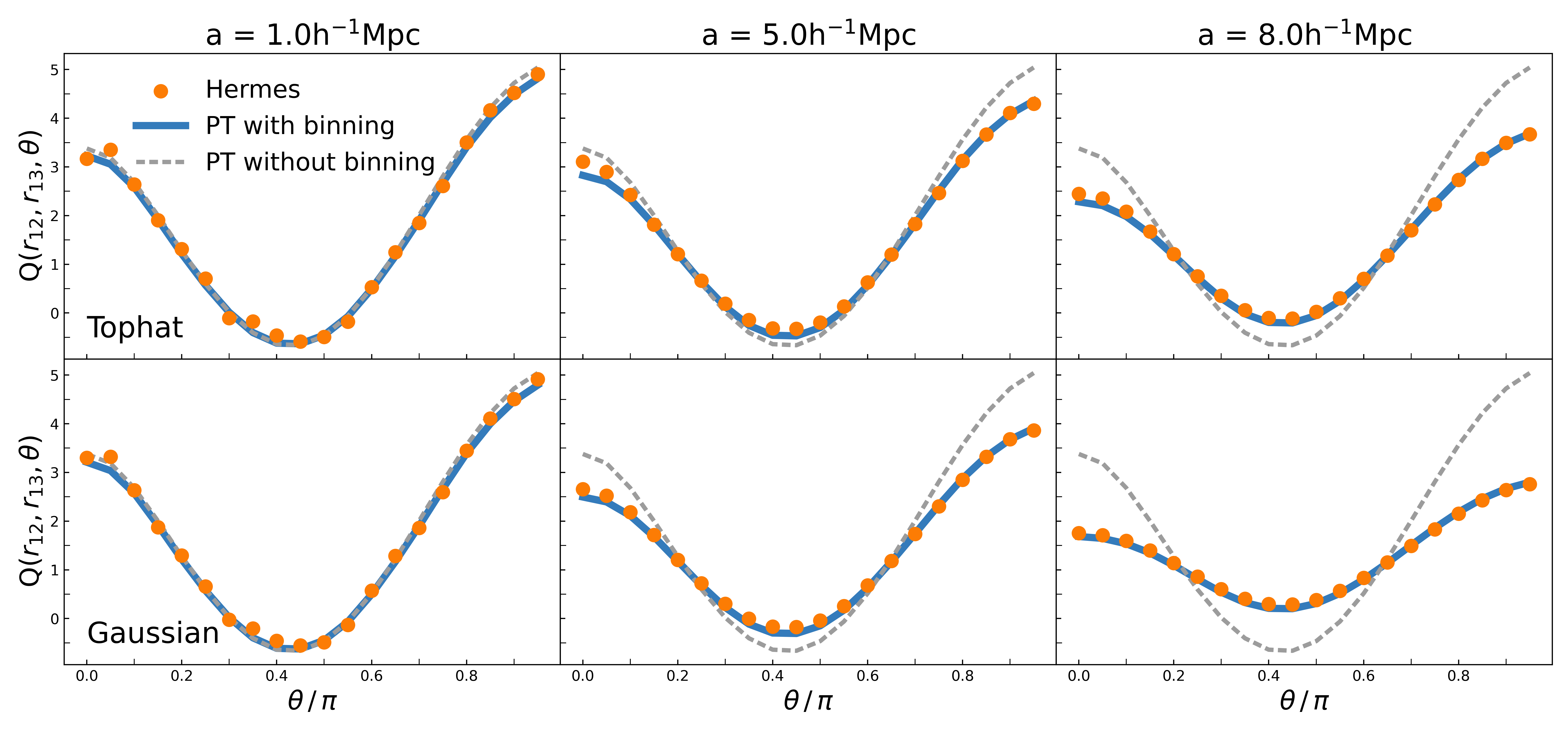}
    \vspace{-0.5cm}
    \caption{Demonstrating the binning effect in the 3PCF measurements: We measure the filtered $Q(r_{12},r_{13},\theta)$ using the spherical tophat (upper) and spherical Gaussian filters (lower) in the MDPL2 dark matter distribution, the triangle configurations are specified by two side lengths of $(r_{12},r_{13}) = (20,40) h^{-1}$Mpc with varying angles $\theta$ between them. The plots from left to right present $Q$ as a function of angle $\theta$ with a tophat filtered radius of 1, 5, 8 $h^{-1}$Mpc respectively, and correspondingly, the Gaussian radius of 0.64, 3.22, 5.14$h^{-1}$Mpc by equal filtered masses. The blue solid line is the binning-corrected theoretical predictions based on tree-level perturbation theory in the $\Lambda$CDM model consistent with the MDPL2 cosmology; in comparison, the black dashed lines without accounting for the binning effect are also plotted.}
    \label{fig:mdpl2 3pcf}
\end{figure*}

In the 3PCF measurements, binning is necessarily introduced to allow for the triplet counting of objects. Alternative to the current binning scheme based on the parameterisation of the 3PCF. we perform triplet-counting by placing a small spherical volume at each vertex of triangles, and the radius of spheres can be, in principle, independent of the triangle configuration. Theoretically, it is equivalent to producing a filtered density field by the spherical top-hat window function, that is, $\delta_W = W\circ \delta$, and 
\begin{equation}\label{eq:3PCF-average}
\zeta_W({\bf r}_1, {\bf r}_2) = \frac{1}{V}\int_V d^3{\bf s}\langle \delta_W({\bf s})\delta_W({\bf s}+{\bf r}_1)\delta_W({\bf s}+{\bf r}_2) \rangle_c ,
\end{equation}
where the ensemble mean is converted to an average over the sampling volume under the ergodic assumption. Given the density field, Eq.~\eqref{eq:3PCF-average} can be evaluated using the Monte-Carlo integration. To minimize shot noise statistically, we employ the edge-corrected estimator to measure the 3PCF by \cite{SzapudiSzalay1998}, 
\begin{equation}
{\zeta} = \frac{(D-R)^3}{R^3} \equiv \frac{\langle(D_1-R_1)(D_2-R_2)(D_3-R_3)\rangle}{\langle R_1R_2R_3 \rangle} ,
\label{eq:SSestimator}
\end{equation}
where the triplet averages are taken over the random sampling points within the entire survey volume. 

Instead of random sampling within the survey volume, an alternative and more efficient method to evaluate the triplet count based on volume averaging is to go through only data points. In this approach, similar to the current way of defining conditional cumulants (e.g. \citealt{Szapudi2009}), triplet counting is performed by using one unfiltered field at a data point and two filtered fields at the other two endpoints. Following this 2-sphere binning scheme, we have, e.g. 
\begin{equation}\label{eq:DDD-sum12}
  DDD = \sum_{i=1}^{N_p} n_W({\bf r}_i+{\bf r}_1) n_W({\bf r}_i+{\bf r}_2)  ,
\end{equation}
where the summation runs over the $N_p$ data points $\{{\bf r}_i, i=1...N_p\}$. Given that $n({\bf r}) =\sum_i w_i\delta_D^3({\bf r}-{\bf r}_i)$, it is easy to see that Eq.~\eqref{eq:DDD-sum12} can be equivalently expressed in the form of a volume average, 
\begin{equation}
  DDD = \frac{1}{V}\int d^3{\bf s}\, n({\bf s})n_W({\bf s}+{\bf r}_1)n_W({\bf s}+{\bf r}_2) .
\end{equation}
In this formulation, Eq.~\eqref{eq:DDD-sum12} provides a statistical estimation of the 3PCF but with a slightly different definition. This alternative approach connects the summation over data points to the continuous volume integral, offering a highly efficient estimation of the 3PCF while still maintaining consistency with volume averaging principles.

As detailed in Section \ref{sec:3PCF-binning}, for a given specific triangle configuration, our optimised 3PCF algorithm assigns two or three spherical windows to the triangle vertices, which are randomly placed within the survey region. Using the {\tt MRACS} strategy, triplet counts within these spheres can be instantly retrieved with $\mathscr{O}(1)$ computational complexity. The corresponding theoretical prediction of the 3PCF, accounting for the binning effects of spherical tophat and Gaussian windows, has been presented in Section \ref{sec:3PCF-theory}.

In the current study of 3PCF, one is used to measure the reduced 3PCF, also referred to as the $Q$ factor, defined by 
\begin{equation}\label{eq:Q-factor}
    \begin{aligned}
    &Q(r_{12},r_{23},r_{31})=\frac{\zeta({\bf r}_1,{\bf r}_2,{\bf r}_3)}{\zeta_H({\bf r}_1,{\bf r}_2,{\bf r}_3)} \\
    &\zeta_H = \xi(r_{12})\xi(r_{23})+\xi(r_{23})\xi(r_{31})+\xi(r_{31})\xi(r_{12})
    \end{aligned}
\end{equation}
where $r_{ij}=|{\bf r}_i-{\bf r}_j|$ is the distance between ${\bf r}_i$ and ${\bf r}_j$. The factor $Q$ was introduced by \cite{Peebles1975, GrothPeebles1977, peebles1980}, and is originally supposed to be constant within the original hierarchical clustering model. However, this assumption has been challenged by more recent surveys and simulations, which reveal a significant dependence of Q on scale (e.g. \citealt{kulkarni2007three, mcbride2010three}) 

In practical 3PCF measurement, we apply the strategy based on the filtered density field method, proceeding in the following steps: 
\begin{itemize}

    \item Produce reference random samples: to perform clustering analysis on a given catalogue using some statistical estimators, we generate a reference random sample with a number density several times that of the catalogue.

    \item Calculate the scaling coefficients: by projecting onto a multiresolution space spanned by a set of basis functions, we calculate the scaling coefficients of density fields of both survey data and random samples, denoted as $\epsilon_{j{\bf l}}$ and $\epsilon_{j{\bf l}}^R$ respectively. We further yield their difference $\Delta\epsilon_{j{\bf l}} = \epsilon_{j{\bf l}} - \epsilon_{j{\bf l}}^R$, which allow us to reconstruct continuous fields for both $\Delta D=D-R$ and $R$.
    
    \item Sample triangular configurations: randomly sample the three vertices that conform to this triangular configuration in space. For pair-counting, we allocate a volume cell to each vertex, which can be described by a binning window function $W$. In the same way, we can have a bi-linear decomposition of the kernel $W$ in the multiresolution space. 
    
    \item Filter the density fields: convolve $\Delta D=D-R$ and $R$ with the window function $W$ to produce the filtered fields $\tilde{D} = W \circ \Delta D $ and $\tilde{R} = W \circ R$. These convolutions can be efficiently performed in the multiresolution space using the FFT technique.

    \item Estimate the 3PCF: by looping over randomly positioned triangles in space, we can obtain estimations of the 3PCF using the generalised edge-corrected estimator Eq.~\eqref{eq:SSestimator}.   
    
\end{itemize}  

\subsection{3PCF: Numerical test}
\label{sec:3PCF_test}

We first test $Q(r_{12},r_{13},\theta)$ estimated in the MDPL2 dark matter simulation sample against the tree-level cosmological perturbation theory including the binning correction. The results of the triangle configuration $(r_{12},r_{13}) = (20,40)h^{-1}$Mpc and $(20,60)h^{-1}$Mpc are shown in Fig.~\eqref{fig:mdpl2 3pcf}. In the measurement, given the sufficiently high number density of dark matter particles in MDPL2, we generated a reference random sample comprising the same count of dark matter particles and applied the edge-corrected estimator Eq.~\eqref{eq:SSestimator}. In addition, a set of Daubechies scaling functions with genius 4 and dilation scale $J=9$ has been used in the analysis. Note that the spatial resolution at $J=9$ is $1.95h^{-1}$Mpc for the MDPL2 sample. Accordingly, we apply the 3-sphere binning scheme using the spherical tophat with filter radii of 1, 5 and 8 $h^{-1}$Mpc.  

Based on Eq.~\eqref{eq:3PCF-average}, under the assumption of ergodicity, the ensemble average can be transformed into a spatial average. Therefore, in evaluating the average via Monto-Carlo integration, it is essential to randomly sample triangles both in spatial positions and orientations in the entire space at an appropriate sampling rate. For a given triangle configuration, our Monte Carlo sampling involves generating $N_p$ positions for the primary vertex of the triangles, followed by $N_{\text{rot}}$ rotations around each primary vertex, resulting in a total of $N_pN_{\text{rot}}$ triangles. At resolution $J=9$, the density field is modelled on a grid $(2^{J})^3 = 512^3$, so we can set $N_p=400^3$. Upon the convergence test, we select $N_{\text{rot}}=2000$ orientations for the angle average. Consequently, the total number of triangles sampled is $1.28\times 10^{11}$. 

To assess the impact of the binning effect, Fig.~\ref{fig:mdpl2 3pcf} also shows the theoretical prediction of $Q$ without the binning correction. On filtered scales below 5$h^{-1}$Mpc, the binning correction introduces only minor changes, except for elongated configurations near $\mu = \cos\theta = \pm 1$. However, if filtered on the typical nonlinear scale of 8$h^{-1}$ Mpc, the binning effect significantly dampens variations in the $Q$ amplitude as a function of angle $\theta$, leading to a notable deviation from the uncorrected theoretical model. As expected, increasing the filtered scale smooths out small-scale noisy fluctuations, resulting in smoother curves in the measurements. Overall, Fig.~\ref{fig:mdpl2 3pcf} demonstrates strong concordance between the theoretical predictions and the simulations. This high degree of agreement is also evident when employing a Gaussian window function. Due to its lack of compact support in real space, the Gaussian filter exerts a more substantial smoothing effect than the spherical top-hat filter, which leads to a more flattened U-shaped characteristic.   

\begin{figure*}
    \centering
    \includegraphics[width=\textwidth]{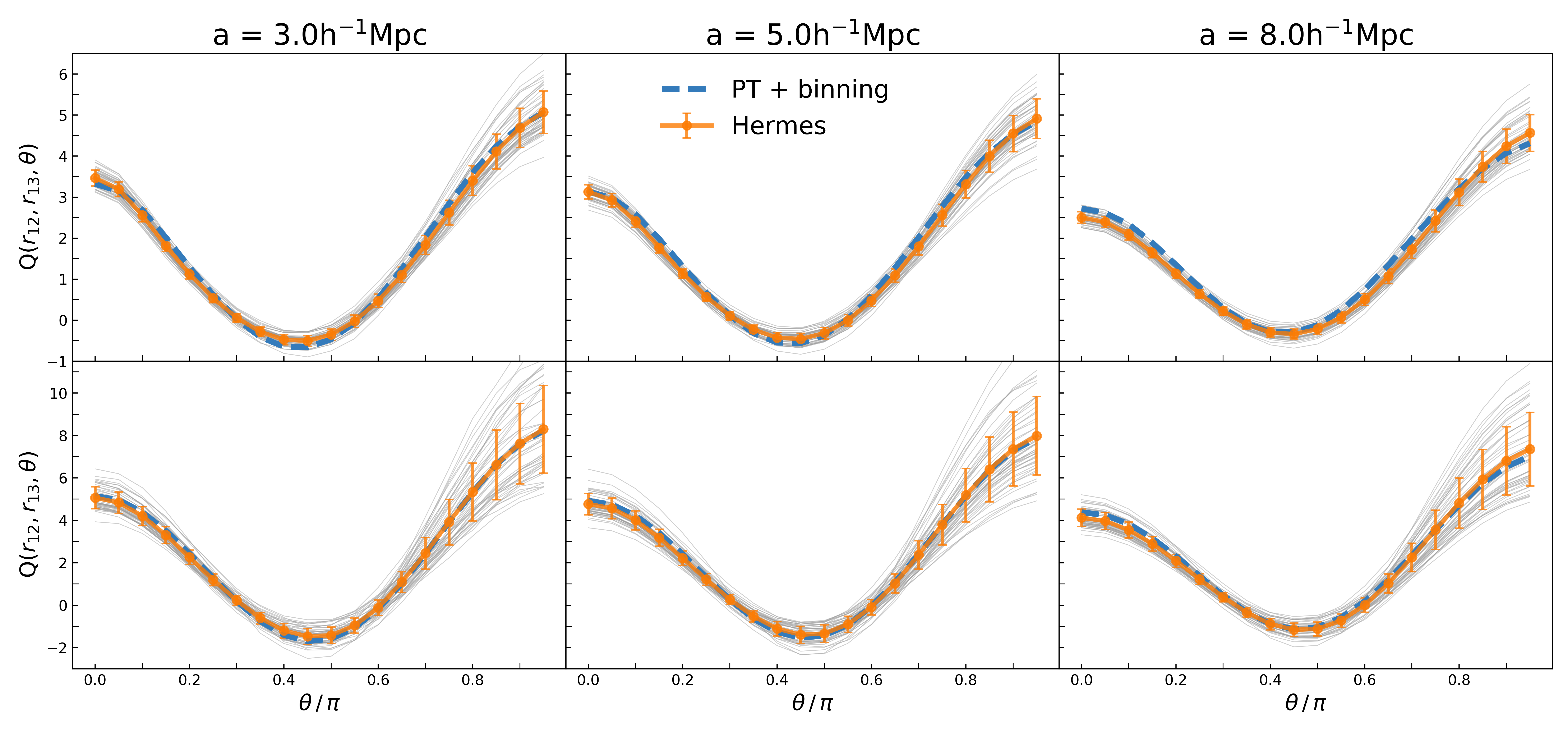}
    \vspace{-0.5cm}
    \caption{The $Q(r_{12},r_{13},\theta)$ as a function of the angle $\theta$, measured in 50 Quijote dark matter catalogues with configuration of $(r_{12},r_{13}) = (20,40)$ and $(20,60)h^{-1}$Mpc, are shown in the upper and lower panels, respectively. The filtered radius are set as 3$h^{-1}$Mpc (left), 5$h^{-1}$Mpc(middle) and 8$h^{-1}$Mpc(right). Each thin grey solid line represents the $Q$ measured in one of 50 Quijote dark matter samples, and the black solid line plots the mean value over 50 samples with 1-$\sigma$ error bars, comparing with theoretical predictions in the tree-level perturbation theory, including the binning effect (blue dashed lines).}
    \label{fig:quijote-dm-3pcf}
\end{figure*}

\begin{figure*}
    \centering
    \includegraphics[width=\textwidth]{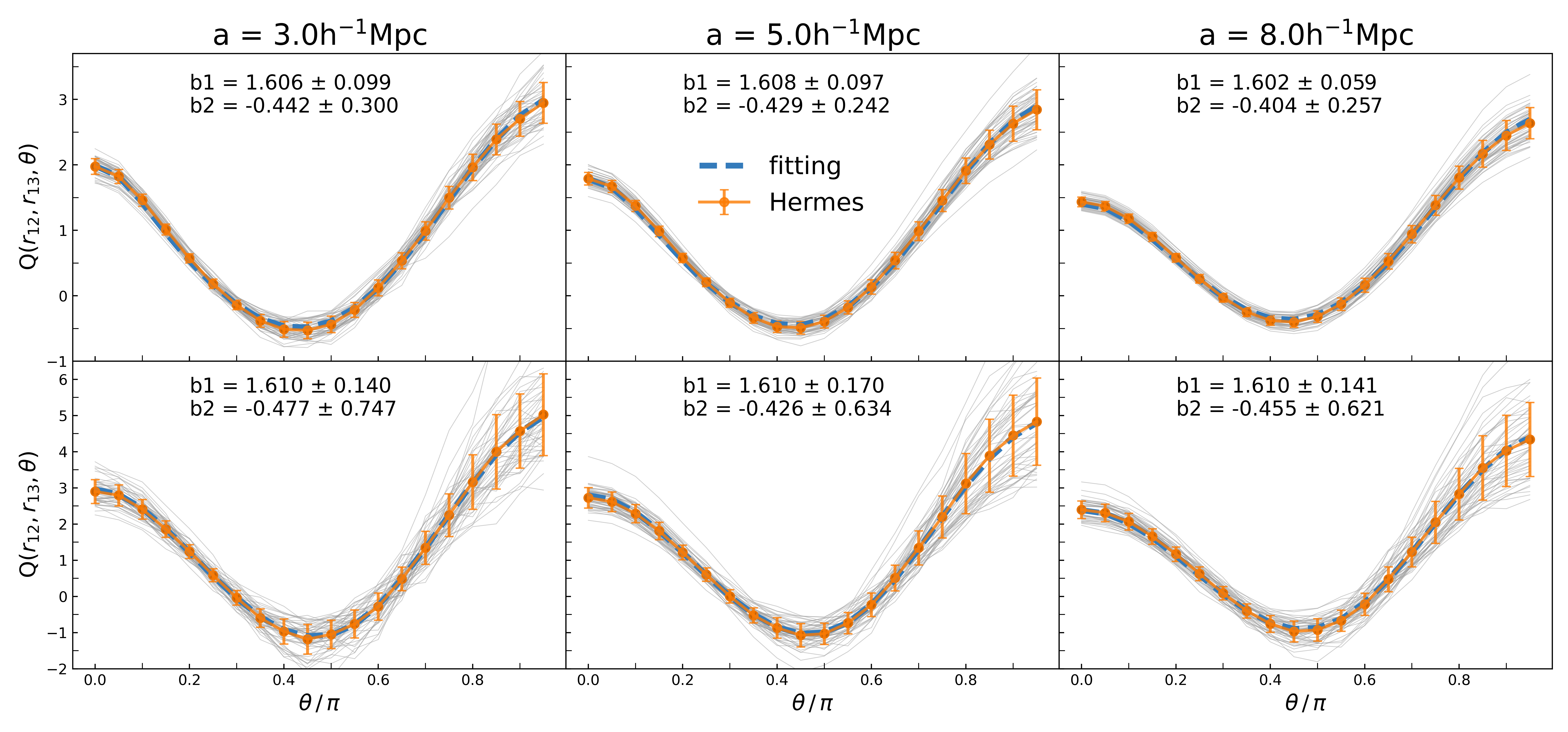}
    \vspace{-0.5cm}
    \caption{Same as Fig.~\ref{fig:quijote-dm-3pcf} but for the 50 halo catalogues extracted from the corresponding dark matter samples. Additionally, the bias parameters $(b_1,b_2)$ obtained by fitting the second-order bias model are listed in the figure, and the corresponding model fitting curves are also plotted (blue dashed line).}
    \label{fig:quijote-halo-3pcf}
\end{figure*}

We performed a 3PCF analysis of 50 realisations of dark matter and their corresponding halo catalogues from the Quijote simulations, which enables us to gauge roughly the statistical errors in the 3PCF measurements. To enhance computational efficiency, we employ a slightly modified binning scheme for triplet counting, specifically the two-sphere binning approach described in Section \ref{sec:3PCF-binning}. In this scheme, the primary vertex of the sampling triangle is iterated over each data particle, and thus the degree of translation $N_p$ is the number of particles in the sample. The remaining configuration freedom of the triangle arises solely from its rotation around the primary in space. This method is particularly effective for sparsely sampled catalogues with a relatively small number of objects, such as halo samples, leading to faster convergence in the 3PCF estimation.

We measure the $Q$ factor as a function of the angle between $r_{12}$ and $r_{13}$ in two configurations of $(r_{12},r_{13}) = (20,40)h^{-1}$Mpc and $(20,60)h^{-1}$Mpc. In addition, we produced the binned density using a spherical top-hat with three different filter radii of $(3, 5, 8)h^{-1}$Mpc. The results are illustrated in Fig.~\ref{fig:quijote-dm-3pcf} and Fig.~\ref{fig:quijote-halo-3pcf}, respectively. As detailed in the appendix, convergence testing reveals that the number of samples required to achieve convergence varies with the scales explored; larger scales demand a higher sampling number. For the Quijote dark matter catalogues, we use $N_{\text{rot}}=200$ only, while for the halo catalogues, we set $N_{\text{rot}}=(5000, 2000, 500)$ for the filtered radii $(3, 5, 8)h^{-1}$Mpc respectively. 

Similarly to 2PCF as described in Section \ref{sec:2PCFtest}, the $Q$ factor measured in dark matter particles is notably smoother than those in halo catalogues, primarily due to the difference in sampling density. Dark matter particles, which are more densely sampled, exhibit less shot noise than sparsely sampled halo catalogues, where increased shot noise on small scales leads to more significant variability in $Q$. Additionally, for a given matter component, the shape of $Q$ tends to become smoother as the filter radius increases. However, as shown in both Fig.~\ref{fig:quijote-dm-3pcf} and Fig.~\ref{fig:quijote-halo-3pcf}, the 1-$\sigma$ variance does not decrease significantly even for the largest filter radii adopted of $8h^{-1}$ Mpc, suggesting that statistical errors remain relatively stable regardless of the filtered radius, but as expected, increases markedly with measured scales due to cosmic variance.

Halos are biased tracers of underlying dark matter distribution. The local relation between the density contrast of halos and dark matter can be expressed by a Taylor series expansion \citep{Fry1993}. Up to the second order, this relation can be written as, 
\begin{equation}\label{eq:fry-biasmodel}
\delta_{\mathrm{h}}(\boldsymbol{x},R)=b_1\delta_{\mathrm{m}}(\boldsymbol{x},R)+\frac{b_2}{2}\left(\delta_{\mathrm{m}}^2(\boldsymbol{x},R)-\left\langle\delta_{\mathrm{m}}^2({\bf x},R)\right\rangle\right),
\end{equation}
in which both the density contrasts $\delta_h$ and $\delta_m$ are smoothed on a given scale $R$ (\citealt{Manera2011,buchalter1999power}). Eq.(\ref{eq:fry-biasmodel}) gives the simplest deterministic nonlinear bias model and neglects the nonlocal contributions such as those arising from the tidal force in inhomogeneous matter fields (e.g., \citealt{chan2012gravity, Baldauf2012-nonlocalbias, Desjacques2018} and references therein). Using the bias model prescribed by Eq.(\ref{eq:fry-biasmodel}), we can relate the Q factor of halos to that of dark matter following from Eq.(\ref{eq:Q-factor}),  
\begin{equation}\label{eq:Q-biasmodel}
    Q_h = \frac{1}{b_1}\Bigl(Q_m+\frac{b_2}{b_1}\Bigr) 
\end{equation}
By combining the $Q$ measurements from dark matter and halo samples in the Quijote simulation, as shown in Fig.~\ref{fig:quijote-dm-3pcf} and Fig.~\ref{fig:quijote-halo-3pcf}, we applied the bias model described in Eq.~\eqref{eq:Q-biasmodel} to fit the bias parameters using the least squares method. The mean values and variances of $b_1$ and $b_2 $ at different bin sizes are provided in Fig.~\ref{fig:quijote-halo-3pcf} in the two triangle configurations $(r_{12}, r_{13}) = (20, 40)\, h^{-1}\text{Mpc}$ and $(20, 60)\, h^{-1}\text{Mpc}$. The results indicate that the bias parameters remain nearly constant across the scales of $20 \sim 80\, h^{-1}\text{Mpc}$. Compared to the results obtained from the 2PCF measurements shown in Fig.~\ref{fig:quijote 2pcf}, the 3PCF measurements yield approximately a $20\%$  overestimation of the linear bias parameter. This result is consistent with the finding given in \citet{Bel2015-nonlocal}, where a $20\%-30\%$ overestimation of the linear bias was observed in 3PCF when the nonlocal bias was ignored. Although this paper does not focus on the halo bias model, we provide this demonstration to replicate the previous findings. 

%%%%%%%%%%%%%%%%%%%%%%%%%%%%%%%%%%%%%%%%%%%%%%%%%%%%%%%%%%%%%%%%%%%%%%%

\section{Concluding Remarks}

This paper presents a novel perspective on the correlation function in the clustering analysis of the large-scale structure of the universe. Based on this view and incorporating the {\tt MRCAS} algorithm within the framework of multiresolution analysis, a highly efficient solution is proposed to tackle the intensive computation in cosmic statistics. The major results and the concluding remarks are summarised as follows.  

(1) According to the working definition of 2PCF based on various statistical estimators, we recognise that pair counting in bins is mathematically equivalent to evaluating CIC, which can be accomplished by convolving the density field with a window function. This insight yields an in-situ expression of 2PCF. Namely, the ex-situ 2-point autocorrelation function can be understood as an in-situ cross-correlation between the original density field and its filtered one. The filter applied depends on the binning scheme in the pair-counting process. In this approach, the binning effect can be easily incorporated into theoretical models without leading to any ambiguities while comparing theoretical predictions with mock or observational data. It is emphasised here that sharp edged filters are necessary for pair counting in a discrete point sample; however, our approach is based on the reconstructed continuous density field and extends the traditional concept of arithmetic pair counting, offering a generalised view of 2PCF with an arbitrarily chosen filter. Of importance, various cosmological applications can benefit from the flexibility in the filter selection, especially with the availability of a rich variety of non-sharp edged filters to adapt to specific purpose. For example, in the forthcoming paper (Li et al., in preparation), we introduced two axial symmetric window functions - disk and cylinder shell - to extract cosmological information in redshift space, and the results show that it can significantly improve the constraints on cosmological parameter estimation. Thus, it can be reasonably concluded that an optimal filter can be specifically designed so that the maximum information content encoded in the survey data can be extracted, although much work remains to be done in this respect.

(2) Based on the in situ expression of 2PCF, the {\tt MRACS} scheme offers a highly efficient algorithm to calculate 2PCF, as detailed in \cite{Feng2007} and Section \ref{sec:MRACS}. The core principle of the {\tt MRACS} scheme is to reconstruct density fields using a set of basis functions within the framework of multiresolution analysis. This approach compresses the (filtered or binned) density field into a data vector comprising scaling coefficients obtained by decomposition on a functional basis, allowing the evaluation of 2PCF to be reduced to a scalar product between two vectors. The computational complexity of this algorithm comes only from performing FFT in convolution. It should be noted that the algorithm itself remains independent of the number of objects, the weighting method, and the binning scheme that includes the geometric shape and size of the filter, establishing it as the fastest algorithm available to date.

(3) When extending to NPCF, it is crucial to realise that the binning scheme may not necessarily depend on the specific N-point polyhedral configurations. In particular, we focus on a fast algorithm in the application to 3PCF. In this approach, triplet counting is practically achieved by assigning two or three spherical filters, tophat or Gaussian, to three vertices of triangles with a given shape. Obviously, this algorithm leads to a 3PCF of the filtered density field. Theoretically, it requires a modified version of the relation between 3PCF and bispectrum. Section \ref{sec:3PCF-theory} gives an analytical expression for 3PCF in multipole decomposition, which accounts for the binning effect introduced by tophat or Gaussian window functions. The numerical tests show excellent agreement with the theoretical predictions based on tree-level perturbation theory in the linear regime.  

This study presents a conceptual extension of traditional N-point clustering statistics. The numerical technique developed accordingly can be applied to variants of N-point statistics, including various weighted or marked auto/cross-correlation functions, multi-point probability distribution functions for CIC, etc. With the advent of extensive photometric and spectroscopic data from current and upcoming sky surveys such as DESI and Euclid, there is an increasing challenge in cosmological data analysis. Our highly efficient algorithm will facilitate a variety of clustering analysis in the large configuration space, unveiling the wealth of encoded cosmological information, e.g., improving the measurement accuracy of cosmological parameters (e.g., \citealt{Gagrani2017, Agarwal2021, AlamEtal2021, Gualdi2021, Samushia2021, Novell-Masot2023}), detecting primordial non-Gaussianity (e.g., \citealt{Komatsu2001, Fergusson2009, chen2010, Desjacques2010, Scoccimarro2012, Biagetti2019, Meerburg2019, Achucarro2022} and references therein), and general relativity effect on cosmic scales (e.g., \citealt{Bonvin2011, Bonvin2014, Bonvin2016, Gaztanaga2017, Tansella_2018, Dio_2019, Beutler_2020, Maartens_2021, Saga2022}). 

\section*{Acknowledgements}

The CosmoSim database used in this paper is a service by the Leibniz-Institute for Astrophysics Potsdam (AIP). The MultiDark database was developed in cooperation with the Spanish MultiDark Consolider Project CSD2009-00064. The authors gratefully acknowledge the Gauss Centre for Supercomputing e.V. (\url{www.gauss-centre.eu}) and the Partnership for Advanced Supercomputing in Europe (PRACE, \url{www.prace-ri.eu}) for funding the MultiDark simulation project by providing computing time on the GCS Supercomputer SuperMUC at the Leibniz Supercomputing Centre (LRZ, \url{www.lrz.de}). FLL is supported by the National Key R\&D Program of China through grant 2020YFC2201400 and the NSFC Key Program through the grants 11733010 and 11333008. JP appreciate the support by NSFC through grant No. 12273049. ZH is supported by the NFSC general program (grant no. 12073088), the National key R\&D Program of China (grant no. 2020YFC2201600), and the National SKA Program of China no. 2020SKA0110402. YC acknowledges the support of
the Royal Society through a University Research Fellowship. For the purpose of open access, the author has applied a Creative Commons Attribution (CC BY) license to any Author Accepted Manuscript version arising from this submission.

%%%%%%%%%%%%%%%%%%%%%%%%%%%%%%%%%%%%%%%%%%%%%%%%%%
\section*{Data Availability}

All simulation data analysed in this paper are available through the MultiDark database website \url{https://www.cosmosim.org} and the Quojote website \url{https://quijote-simulations.readthedocs.io/en/latest/}. The authors agree to make data supporting the results or analyses presented in their paper available upon reasonable request. The {\tt Hermes} toolkit will be publicly available on GitHub together with the release of the relevant code paper soon (Feng et al., in preparation). An early test version of {\tt Hermes} is also available from the corresponding author upon special request. 



\bibliographystyle{mnras}
\bibliography{FilteredCF} % if your bibtex file is called example.bib

\begin{thebibliography}{}
\makeatletter
\relax
\def\mn@urlcharsother{\let\do\@makeother \do\$\do\&\do\#\do\^\do\_\do\%\do\~}
\def\mn@doi{\begingroup\mn@urlcharsother \@ifnextchar [ {\mn@doi@} {\mn@doi@[]}}
\def\mn@doi@[#1]#2{\def\@tempa{#1}\ifx\@tempa\@empty \href {http://dx.doi.org/#2} {doi:#2}\else \href {http://dx.doi.org/#2} {#1}\fi \endgroup}
\def\mn@eprint#1#2{\mn@eprint@#1:#2::\@nil}
\def\mn@eprint@arXiv#1{\href {http://arxiv.org/abs/#1} {{\tt arXiv:#1}}}
\def\mn@eprint@dblp#1{\href {http://dblp.uni-trier.de/rec/bibtex/#1.xml} {dblp:#1}}
\def\mn@eprint@#1:#2:#3:#4\@nil{\def\@tempa {#1}\def\@tempb {#2}\def\@tempc {#3}\ifx \@tempc \@empty \let \@tempc \@tempb \let \@tempb \@tempa \fi \ifx \@tempb \@empty \def\@tempb {arXiv}\fi \@ifundefined {mn@eprint@\@tempb}{\@tempb:\@tempc}{\expandafter \expandafter \csname mn@eprint@\@tempb\endcsname \expandafter{\@tempc}}}

\bibitem[\protect\citeauthoryear{{Ach{\'u}carro} et~al.,}{{Ach{\'u}carro} et~al.}{2022}]{Achucarro2022}
{Ach{\'u}carro} A.,  et~al., 2022, \mn@doi [arXiv e-prints] {10.48550/arXiv.2203.08128}, \href {https://ui.adsabs.harvard.edu/abs/2022arXiv220308128A} {p. arXiv:2203.08128}

\bibitem[\protect\citeauthoryear{{Agarwal}, {Desjacques}, {Jeong}  \& {Schmidt}}{{Agarwal} et~al.}{2021}]{Agarwal2021}
{Agarwal} N.,  {Desjacques} V.,  {Jeong} D.,   {Schmidt} F.,  2021, \mn@doi [\jcap] {10.1088/1475-7516/2021/03/021}, \href {https://ui.adsabs.harvard.edu/abs/2021\jcap...03..021A} {2021, 021}

\bibitem[\protect\citeauthoryear{{Akeson} et~al.,}{{Akeson} et~al.}{2019}]{Akeson2019}
{Akeson} R.,  et~al., 2019, \mn@doi [arXiv e-prints] {10.48550/arXiv.1902.05569}, \href {https://ui.adsabs.harvard.edu/abs/2019arXiv190205569A} {p. arXiv:1902.05569}

\bibitem[\protect\citeauthoryear{Alam et~al.,}{Alam et~al.}{2017}]{Alam2017}
Alam S.,  et~al., 2017, \mn@doi [\mnras] {10.1093/\mnras/stx721}, 470, 2617

\bibitem[\protect\citeauthoryear{{Alam} et~al.,}{{Alam} et~al.}{2021a}]{Alam2021_SDSS}
{Alam} S.,  et~al., 2021a, \mn@doi [\prd] {10.1103/PhysRevD.103.083533}, \href {https://ui.adsabs.harvard.edu/abs/2021PhRvD.103h3533A} {103, 083533}

\bibitem[\protect\citeauthoryear{{Alam} et~al.,}{{Alam} et~al.}{2021b}]{Alam2021_DESI}
{Alam} S.,  et~al., 2021b, \mn@doi [\jcap] {10.1088/1475-7516/2021/11/050}, \href {https://ui.adsabs.harvard.edu/abs/2021\jcap...11..050A} {2021, 050}

\bibitem[\protect\citeauthoryear{{Alam} et~al.,}{{Alam} et~al.}{2021c}]{AlamEtal2021}
{Alam} S.,  et~al., 2021c, \mn@doi [\jcap] {10.1088/1475-7516/2021/11/050}, \href {https://ui.adsabs.harvard.edu/abs/2021\jcap...11..050A} {2021, 050}

\bibitem[\protect\citeauthoryear{Anderson et~al.,}{Anderson et~al.}{2014}]{Anderson2014}
Anderson L.,  et~al., 2014, \mn@doi [\mnras] {10.1093/\mnras/stu523}, 441, 24

\bibitem[\protect\citeauthoryear{Armijo, Cai, Padilla, Li  \& Peacock}{Armijo et~al.}{2018}]{Armijo2018}
Armijo J.,  Cai Y.-C.,  Padilla N.,  Li B.,   Peacock J.~A.,  2018, \mn@doi [\mnras] {10.1093/\mnras/sty1335}, 478, 3627

\bibitem[\protect\citeauthoryear{{Bailoni}, {Spurio Mancini}  \& {Amendola}}{{Bailoni} et~al.}{2017}]{Bailoni2017}
{Bailoni} A.,  {Spurio Mancini} A.,   {Amendola} L.,  2017, \mn@doi [\mnras] {10.1093/mnras/stx1209}, \href {https://ui.adsabs.harvard.edu/abs/2017MNRAS.470..688B} {470, 688}

\bibitem[\protect\citeauthoryear{Baldauf, Seljak, Desjacques  \& McDonald}{Baldauf et~al.}{2012}]{Baldauf2012-nonlocalbias}
Baldauf T.,  Seljak U. c.~v.,  Desjacques V.,   McDonald P.,  2012, \mn@doi [\prd] {10.1103/PhysRevD.86.083540}, 86, 083540

\bibitem[\protect\citeauthoryear{{Baxter} \& {Rozo}}{{Baxter} \& {Rozo}}{2013}]{BaxterRozo2013}
{Baxter} E.~J.,  {Rozo} E.,  2013, \mn@doi [\apj] {10.1088/0004-637X/779/1/62}, \href {https://ui.adsabs.harvard.edu/abs/2013ApJ...779...62B} {779, 62}

\bibitem[\protect\citeauthoryear{Bel, Hoffmann  \& Gaztañaga}{Bel et~al.}{2015}]{Bel2015-nonlocal}
Bel J.,  Hoffmann K.,   Gaztañaga E.,  2015, \mn@doi [\mnras] {10.1093/\mnras/stv1600}, 453, 259

\bibitem[\protect\citeauthoryear{{Bernardeau}, {Colombi}, {Gazta{\~n}aga}  \& {Scoccimarro}}{{Bernardeau} et~al.}{2002}]{Bernardeau2002}
{Bernardeau} F.,  {Colombi} S.,  {Gazta{\~n}aga} E.,   {Scoccimarro} R.,  2002, \mn@doi [\physrep] {10.1016/S0370-1573(02)00135-7}, \href {https://ui.adsabs.harvard.edu/abs/2002PhR...367....1B} {367, 1}

\bibitem[\protect\citeauthoryear{Beutler \& Dio}{Beutler \& Dio}{2020}]{Beutler_2020}
Beutler F.,  Dio E.~D.,  2020, \mn@doi [\jcap] {10.1088/1475-7516/2020/07/048}, 2020, 048

\bibitem[\protect\citeauthoryear{Beutler et~al.,}{Beutler et~al.}{2011}]{Beutler2011}
Beutler F.,  et~al., 2011, \mn@doi [\mnras] {10.1111/j.1365-2966.2011.19250.x}, 416, 3017

\bibitem[\protect\citeauthoryear{{Biagetti}}{{Biagetti}}{2019}]{Biagetti2019}
{Biagetti} M.,  2019, \mn@doi [Galaxies] {10.3390/galaxies7030071}, \href {https://ui.adsabs.harvard.edu/abs/2019Galax...7...71B} {7, 71}

\bibitem[\protect\citeauthoryear{Blake et~al.,}{Blake et~al.}{2011}]{Blake2011}
Blake C.,  et~al., 2011, \mn@doi [\mnras] {10.1111/j.1365-2966.2011.19592.x}, 418, 1707

\bibitem[\protect\citeauthoryear{Bonvin \& Durrer}{Bonvin \& Durrer}{2011}]{Bonvin2011}
Bonvin C.,  Durrer R.,  2011, \mn@doi [\prd] {10.1103/PhysRevD.84.063505}, 84, 063505

\bibitem[\protect\citeauthoryear{Bonvin, Hui  \& Gazta\~naga}{Bonvin et~al.}{2014}]{Bonvin2014}
Bonvin C.,  Hui L.,   Gazta\~naga E.,  2014, \mn@doi [\prd] {10.1103/PhysRevD.89.083535}, 89, 083535

\bibitem[\protect\citeauthoryear{{Bonvin}, {Hui}  \& {Gaztanaga}}{{Bonvin} et~al.}{2016a}]{buchalter1999power}
{Bonvin} C.,  {Hui} L.,   {Gaztanaga} E.,  2016a, \mn@doi [\jcap] {10.1088/1475-7516/2016/08/021}, \href {https://ui.adsabs.harvard.edu/abs/2016JCAP...08..021B} {2016, 021}

\bibitem[\protect\citeauthoryear{{Bonvin}, {Hui}  \& {Gaztanaga}}{{Bonvin} et~al.}{2016b}]{Bonvin2016}
{Bonvin} C.,  {Hui} L.,   {Gaztanaga} E.,  2016b, \mn@doi [\jcap] {10.1088/1475-7516/2016/08/021}, \href {https://ui.adsabs.harvard.edu/abs/2016JCAP...08..021B} {2016, 021}

\bibitem[\protect\citeauthoryear{{Breton} \& {de la Torre}}{{Breton} \& {de la Torre}}{2021}]{breton2021fast}
{Breton} M.-A.,  {de la Torre} S.,  2021, \mn@doi [\aap] {10.1051/0004-6361/202039603}, \href {https://ui.adsabs.harvard.edu/abs/2021A&A...646A..40B} {646, A40}

\bibitem[\protect\citeauthoryear{{Brown}, {Mishtaku}  \& {Demina}}{{Brown} et~al.}{2022}]{BrownEtal2022}
{Brown} Z.,  {Mishtaku} G.,   {Demina} R.,  2022, \mn@doi [\aap] {10.1051/0004-6361/202141917}, \href {https://ui.adsabs.harvard.edu/abs/2022A&A...667A.129B} {667, A129}

\bibitem[\protect\citeauthoryear{{Chan}, {Scoccimarro}  \& {Sheth}}{{Chan} et~al.}{2012}]{chan2012gravity}
{Chan} K.~C.,  {Scoccimarro} R.,   {Sheth} R.~K.,  2012, \mn@doi [\prd] {10.1103/PhysRevD.85.083509}, \href {https://ui.adsabs.harvard.edu/abs/2012PhRvD..85h3509C} {85, 083509}

\bibitem[\protect\citeauthoryear{{Chen}}{{Chen}}{2010}]{chen2010}
{Chen} X.,  2010, \mn@doi [Advances in Astronomy] {10.1155/2010/638979}, \href {https://ui.adsabs.harvard.edu/abs/2010AdAst2010E..72C} {2010, 638979}

\bibitem[\protect\citeauthoryear{Cohn}{Cohn}{2006}]{Cohn2006}
Cohn J.,  2006, \mn@doi [New Astronomy] {10.1016/j.newast.2005.08.002}, 11, 226

\bibitem[\protect\citeauthoryear{{Cole} et~al.,}{{Cole} et~al.}{2005}]{cole2005}
{Cole} S.,  et~al., 2005, \mn@doi [\mnras] {10.1111/j.1365-2966.2005.09318.x}, \href {https://ui.adsabs.harvard.edu/abs/2005MNRAS.362..505C} {362, 505}

\bibitem[\protect\citeauthoryear{Crocce \& Scoccimarro}{Crocce \& Scoccimarro}{2006}]{Crocce2006}
Crocce M.,  Scoccimarro R.,  2006, \mn@doi [\prd] {10.1103/PhysRevD.73.063520}, 73, 063520

\bibitem[\protect\citeauthoryear{Crocce \& Scoccimarro}{Crocce \& Scoccimarro}{2008}]{Crocce2008}
Crocce M.,  Scoccimarro R.,  2008, \mn@doi [\prd] {10.1103/PhysRevD.77.023533}, 77, 023533

\bibitem[\protect\citeauthoryear{{Cui}, {Liu}, {Yang}, {Wang}, {Feng}  \& {Springel}}{{Cui} et~al.}{2008}]{Cui2008}
{Cui} W.,  {Liu} L.,  {Yang} X.,  {Wang} Y.,  {Feng} L.,   {Springel} V.,  2008, \mn@doi [\apj] {10.1086/592079}, \href {https://ui.adsabs.harvard.edu/abs/2008\apj...687..738C} {687, 738}

\bibitem[\protect\citeauthoryear{{DESI Collaboration} et~al.,}{{DESI Collaboration} et~al.}{2016}]{DESI2016}
{DESI Collaboration} et~al., 2016, \mn@doi [arXiv e-prints] {10.48550/arXiv.1611.00036}, \href {https://ui.adsabs.harvard.edu/abs/2016arXiv161100036D} {p. arXiv:1611.00036}

\bibitem[\protect\citeauthoryear{{DESI Collaboration} et~al.,}{{DESI Collaboration} et~al.}{2024}]{DESI_BAO2024}
{DESI Collaboration} et~al., 2024, \mn@doi [arXiv e-prints] {10.48550/arXiv.2404.03000}, \href {https://ui.adsabs.harvard.edu/abs/2024arXiv240403000D} {p. arXiv:2404.03000}

\bibitem[\protect\citeauthoryear{{D{\'a}vila-Kurb{\'a}n}, {S{\'a}nchez}, {Lares}  \& {Ruiz}}{{D{\'a}vila-Kurb{\'a}n} et~al.}{2021}]{DavilaEtal2021}
{D{\'a}vila-Kurb{\'a}n} F.,  {S{\'a}nchez} A.~G.,  {Lares} M.,   {Ruiz} A.~N.,  2021, \mn@doi [\mnras] {10.1093/\mnras/stab1622}, \href {https://ui.adsabs.harvard.edu/abs/2021\mnras.506.4667D} {506, 4667}

\bibitem[\protect\citeauthoryear{{Demina}, {Cheong}, {BenZvi}  \& {Hindrichs}}{{Demina} et~al.}{2018}]{DeminaEtal2018}
{Demina} R.,  {Cheong} S.,  {BenZvi} S.,   {Hindrichs} O.,  2018, \mn@doi [\mnras] {10.1093/\mnras/sty1812}, \href {https://ui.adsabs.harvard.edu/abs/2018\mnras.480...49D} {480, 49}

\bibitem[\protect\citeauthoryear{Desjacques \& Seljak}{Desjacques \& Seljak}{2010}]{Desjacques2010}
Desjacques V.,  Seljak U.,  2010, \mn@doi [Classical and Quantum Gravity] {10.1088/0264-9381/27/12/124011}, 27, 124011

\bibitem[\protect\citeauthoryear{Desjacques, Jeong  \& Schmidt}{Desjacques et~al.}{2018}]{Desjacques2018}
Desjacques V.,  Jeong D.,   Schmidt F.,  2018, \mn@doi [Physics Reports] {10.1016/j.physrep.2017.12.002}, 733, 1

\bibitem[\protect\citeauthoryear{Dio \& Seljak}{Dio \& Seljak}{2019}]{Dio_2019}
Dio E.~D.,  Seljak U.,  2019, \mn@doi [\jcap] {10.1088/1475-7516/2019/04/050}, 2019, 050

\bibitem[\protect\citeauthoryear{{Dirac}}{{Dirac}}{1947}]{dirac1949principles}
{Dirac} P.~A.~M.,  1947, {The principles of quantum mechanics}

\bibitem[\protect\citeauthoryear{Donoso}{Donoso}{2019}]{Donoso2019}
Donoso E.,  2019, \mn@doi [\mnras] {10.1093/\mnras/stz1469}, 487, 2824

\bibitem[\protect\citeauthoryear{Eisenstein et~al.,}{Eisenstein et~al.}{2005a}]{Eisenstein_2005}
Eisenstein D.~J.,  et~al., 2005a, \mn@doi [\apj] {10.1086/466512}, 633, 560

\bibitem[\protect\citeauthoryear{{Eisenstein} et~al.,}{{Eisenstein} et~al.}{2005b}]{Eisenstein2005}
{Eisenstein} D.~J.,  et~al., 2005b, \mn@doi [\apj] {10.1086/466512}, \href {https://ui.adsabs.harvard.edu/abs/2005\apj...633..560E} {633, 560}

\bibitem[\protect\citeauthoryear{{Eisenstein} et~al.,}{{Eisenstein} et~al.}{2005c}]{eisenstein2005detection}
{Eisenstein} D.~J.,  et~al., 2005c, \mn@doi [\apj] {10.1086/466512}, \href {https://ui.adsabs.harvard.edu/abs/2005ApJ...633..560E} {633, 560}

\bibitem[\protect\citeauthoryear{{Eisenstein}, {Seo}  \& {White}}{{Eisenstein} et~al.}{2007}]{eisenstein2007robustness}
{Eisenstein} D.~J.,  {Seo} H.-J.,   {White} M.,  2007, \mn@doi [\apj] {10.1086/518755}, \href {https://ui.adsabs.harvard.edu/abs/2007ApJ...664..660E} {664, 660}

\bibitem[\protect\citeauthoryear{Feng}{Feng}{2007}]{Feng2007}
Feng L.-L.,  2007, \mn@doi [\apj] {10.1086/511024}, 658, 25

\bibitem[\protect\citeauthoryear{Feng \& Fang}{Feng \& Fang}{2000}]{Feng2000}
Feng L.-L.,  Fang L.-Z.,  2000, \mn@doi [\apj] {10.1086/308874}, 535, 519

\bibitem[\protect\citeauthoryear{Fergusson \& Shellard}{Fergusson \& Shellard}{2009}]{Fergusson2009}
Fergusson J.~R.,  Shellard E. P.~S.,  2009, \mn@doi [\prd] {10.1103/PhysRevD.80.043510}, 80, 043510

\bibitem[\protect\citeauthoryear{{Frieman} \& {Gaztanaga}}{{Frieman} \& {Gaztanaga}}{1994}]{Frieman1994}
{Frieman} J.~A.,  {Gaztanaga} E.,  1994, \mn@doi [\apj] {10.1086/173995}, \href {https://ui.adsabs.harvard.edu/abs/1994\apj...425..392F} {425, 392}

\bibitem[\protect\citeauthoryear{{Fry} \& {Gaztanaga}}{{Fry} \& {Gaztanaga}}{1993}]{Fry1993}
{Fry} J.~N.,  {Gaztanaga} E.,  1993, \mn@doi [\apj] {10.1086/173015}, \href {https://ui.adsabs.harvard.edu/abs/1993\apj...413..447F} {413, 447}

\bibitem[\protect\citeauthoryear{{Gagrani} \& {Samushia}}{{Gagrani} \& {Samushia}}{2017}]{Gagrani2017}
{Gagrani} P.,  {Samushia} L.,  2017, \mn@doi [\mnras] {10.1093/\mnras/stx135}, \href {https://ui.adsabs.harvard.edu/abs/2017\mnras.467..928G} {467, 928}

\bibitem[\protect\citeauthoryear{{Gaztanaga} \& {Frieman}}{{Gaztanaga} \& {Frieman}}{1994}]{Gaztanaga1994}
{Gaztanaga} E.,  {Frieman} J.~A.,  1994, \mn@doi [\apjl] {10.1086/187671}, \href {https://ui.adsabs.harvard.edu/abs/1994\apj...437L..13G} {437, L13}

\bibitem[\protect\citeauthoryear{{Gaztanaga}, {Bonvin}  \& {Hui}}{{Gaztanaga} et~al.}{2017}]{Gaztanaga2017}
{Gaztanaga} E.,  {Bonvin} C.,   {Hui} L.,  2017, \mn@doi [\jcap] {10.1088/1475-7516/2017/01/032}, \href {https://ui.adsabs.harvard.edu/abs/2017\jcap...01..032G} {2017, 032}

\bibitem[\protect\citeauthoryear{Gaztañaga \& Scoccimarro}{Gaztañaga \& Scoccimarro}{2005}]{Gaztanaga2005}
Gaztañaga E.,  Scoccimarro R.,  2005, \mn@doi [\mnras] {10.1111/j.1365-2966.2005.09234.x}, 361, 824

\bibitem[\protect\citeauthoryear{{Gil-Mar{\'\i}n}, {Nore{\~n}a}, {Verde}, {Percival}, {Wagner}, {Manera}  \& {Schneider}}{{Gil-Mar{\'\i}n} et~al.}{2015}]{GilMar2015}
{Gil-Mar{\'\i}n} H.,  {Nore{\~n}a} J.,  {Verde} L.,  {Percival} W.~J.,  {Wagner} C.,  {Manera} M.,   {Schneider} D.~P.,  2015, \mn@doi [\mnras] {10.1093/\mnras/stv961}, \href {https://ui.adsabs.harvard.edu/abs/2015\mnras.451..539G} {451, 539}

\bibitem[\protect\citeauthoryear{{Gil-Mar{\'\i}n}, {Percival}, {Verde}, {Brownstein}, {Chuang}, {Kitaura}, {Rodr{\'\i}guez-Torres}  \& {Olmstead}}{{Gil-Mar{\'\i}n} et~al.}{2017}]{GilMar2017}
{Gil-Mar{\'\i}n} H.,  {Percival} W.~J.,  {Verde} L.,  {Brownstein} J.~R.,  {Chuang} C.-H.,  {Kitaura} F.-S.,  {Rodr{\'\i}guez-Torres} S.~A.,   {Olmstead} M.~D.,  2017, \mn@doi [\mnras] {10.1093/\mnras/stw2679}, \href {https://ui.adsabs.harvard.edu/abs/2017\mnras.465.1757G} {465, 1757}

\bibitem[\protect\citeauthoryear{{Groth} \& {Peebles}}{{Groth} \& {Peebles}}{1977}]{GrothPeebles1977}
{Groth} E.~J.,  {Peebles} P.~J.~E.,  1977, \mn@doi [\apj] {10.1086/155588}, \href {https://ui.adsabs.harvard.edu/abs/1977ApJ...217..385G} {217, 385}

\bibitem[\protect\citeauthoryear{Gualdi, Gil-Marín  \& Verde}{Gualdi et~al.}{2021}]{Gualdi2021}
Gualdi D.,  Gil-Marín H.,   Verde L.,  2021, \mn@doi [\jcap] {10.1088/1475-7516/2021/07/008}, 2021, 008

\bibitem[\protect\citeauthoryear{Guo et~al.,}{Guo et~al.}{2015}]{Guo2015}
Guo H.,  et~al., 2015, \mn@doi [Monthly Notices of the Royal Astronomical Society: Letters] {10.1093/mnrasl/slv020}, 449, L95

\bibitem[\protect\citeauthoryear{{Guth} \& {Pi}}{{Guth} \& {Pi}}{1982}]{guth1982}
{Guth} A.~H.,  {Pi} S.~Y.,  1982, \mn@doi [\prl] {10.1103/PhysRevLett.49.1110}, \href {https://ui.adsabs.harvard.edu/abs/1982PhRvL..49.1110G} {49, 1110}

\bibitem[\protect\citeauthoryear{{Hamilton}}{{Hamilton}}{1988}]{Hamilton1988}
{Hamilton} A.~J.~S.,  1988, \mn@doi [\apjl] {10.1086/185235}, \href {https://ui.adsabs.harvard.edu/abs/1988\apj...331L..59H} {331, L59}

\bibitem[\protect\citeauthoryear{{Hamilton}}{{Hamilton}}{1993}]{Hamilton1993}
{Hamilton} A.~J.~S.,  1993, \mn@doi [\apj] {10.1086/173288}, \href {https://ui.adsabs.harvard.edu/abs/1993\apj...417...19H} {417, 19}

\bibitem[\protect\citeauthoryear{Hand, Feng, Beutler, Li, Modi, Seljak  \& Slepian}{Hand et~al.}{2018}]{Hand2018}
Hand N.,  Feng Y.,  Beutler F.,  Li Y.,  Modi C.,  Seljak U.,   Slepian Z.,  2018, \mn@doi [\aj] {10.3847/1538-3881/aadae0}, 156, 160

\bibitem[\protect\citeauthoryear{{Hawking}}{{Hawking}}{1982}]{hawking1982}
{Hawking} S.~W.,  1982, \mn@doi [Physics Letters B] {10.1016/0370-2693(82)90373-2}, \href {https://ui.adsabs.harvard.edu/abs/1982PhLB..115..295H} {115, 295}

\bibitem[\protect\citeauthoryear{{Hawkins} et~al.,}{{Hawkins} et~al.}{2003}]{HawkinsEtal2003}
{Hawkins} E.,  et~al., 2003, \mn@doi [\mnras] {10.1046/j.1365-2966.2003.07063.x}, \href {https://ui.adsabs.harvard.edu/abs/2003\mnras.346...78H} {346, 78}

\bibitem[\protect\citeauthoryear{{He}}{{He}}{2021}]{he2021fast}
{He} C.-C.,  2021, \mn@doi [\apj] {10.3847/1538-4357/ac1daa}, \href {https://ui.adsabs.harvard.edu/abs/2021ApJ...921...59H} {921, 59}

\bibitem[\protect\citeauthoryear{Hernández-Aguayo, Baugh  \& Li}{Hernández-Aguayo et~al.}{2018}]{Aguayo2018}
Hernández-Aguayo C.,  Baugh C.~M.,   Li B.,  2018, \mn@doi [\mnras] {10.1093/\mnras/sty1822}, 479, 4824

\bibitem[\protect\citeauthoryear{Hildebrandt et~al.,}{Hildebrandt et~al.}{2016}]{Hildebrandt2016}
Hildebrandt H.,  et~al., 2016, \mn@doi [\mnras] {10.1093/\mnras/stw2805}, 465, 1454

\bibitem[\protect\citeauthoryear{Ivanov, Simonović  \& Zaldarriaga}{Ivanov et~al.}{2020}]{Ivanov2020}
Ivanov M.~M.,  Simonović M.,   Zaldarriaga M.,  2020, \mn@doi [\jcap] {10.1088/1475-7516/2020/05/042}, 2020, 042

\bibitem[\protect\citeauthoryear{{Ivezi{\'c}} et~al.,}{{Ivezi{\'c}} et~al.}{2019}]{Ivezic2019}
{Ivezi{\'c}} {\v{Z}}.,  et~al., 2019, \mn@doi [\apj] {10.3847/1538-4357/ab042c}, \href {https://ui.adsabs.harvard.edu/abs/2019\apj...873..111I} {873, 111}

\bibitem[\protect\citeauthoryear{{Jang} \& {Meng Loh}}{{Jang} \& {Meng Loh}}{2017}]{JangJi2017}
{Jang} W.,  {Meng Loh} J.,  2017, \mn@doi [\apj] {10.3847/1538-4357/aa67f5}, \href {https://ui.adsabs.harvard.edu/abs/2017ApJ...839...62J} {839, 62}

\bibitem[\protect\citeauthoryear{{Jing} \& {B{\"o}rner}}{{Jing} \& {B{\"o}rner}}{1998}]{Jing1998}
{Jing} Y.~P.,  {B{\"o}rner} G.,  1998, \mn@doi [\apj] {10.1086/305997}, \href {https://ui.adsabs.harvard.edu/abs/1998\apj...503...37J} {503, 37}

\bibitem[\protect\citeauthoryear{Jing \& Börner}{Jing \& Börner}{2004}]{Jing2004}
Jing Y.~P.,  Börner G.,  2004, \mn@doi [\apj] {10.1086/383343}, 607, 140

\bibitem[\protect\citeauthoryear{Karim, Rezaie, Singh  \& Eisenstein}{Karim et~al.}{2023}]{Karim2023}
Karim T.,  Rezaie M.,  Singh S.,   Eisenstein D.,  2023, \mn@doi [\mnras] {10.1093/\mnras/stad2210}, 525, 311

\bibitem[\protect\citeauthoryear{{Kayo} et~al.,}{{Kayo} et~al.}{2004}]{Kayo2004}
{Kayo} I.,  et~al., 2004, \mn@doi [\pasj] {10.1093/pasj/56.3.415}, \href {https://ui.adsabs.harvard.edu/abs/2004PASJ...56..415K} {56, 415}

\bibitem[\protect\citeauthoryear{{Keih{\"a}nen} et~al.,}{{Keih{\"a}nen} et~al.}{2019}]{keihanen2019estimating}
{Keih{\"a}nen} E.,  et~al., 2019, \mn@doi [\aap] {10.1051/0004-6361/201935828}, \href {https://ui.adsabs.harvard.edu/abs/2019A&A...631A..73K} {631, A73}

\bibitem[\protect\citeauthoryear{{Kerscher}}{{Kerscher}}{2022}]{kerscher2022improving}
{Kerscher} M.,  2022, \mn@doi [\aap] {10.1051/0004-6361/202243632}, \href {https://ui.adsabs.harvard.edu/abs/2022A&A...666A.181K} {666, A181}

\bibitem[\protect\citeauthoryear{{Kerscher}, {Szapudi}  \& {Szalay}}{{Kerscher} et~al.}{2000}]{KerscherEtal2000}
{Kerscher} M.,  {Szapudi} I.,   {Szalay} A.~S.,  2000, \mn@doi [\apjl] {10.1086/312702}, \href {https://ui.adsabs.harvard.edu/abs/2000\apj...535L..13K} {535, L13}

\bibitem[\protect\citeauthoryear{Komatsu \& Spergel}{Komatsu \& Spergel}{2001}]{Komatsu2001}
Komatsu E.,  Spergel D.~N.,  2001, \mn@doi [\prd] {10.1103/PhysRevD.63.063002}, 63, 063002

\bibitem[\protect\citeauthoryear{{Kulkarni}, {Nichol}, {Sheth}, {Seo}, {Eisenstein}  \& {Gray}}{{Kulkarni} et~al.}{2007a}]{KulkarniEtal2007}
{Kulkarni} G.~V.,  {Nichol} R.~C.,  {Sheth} R.~K.,  {Seo} H.-J.,  {Eisenstein} D.~J.,   {Gray} A.,  2007a, \mn@doi [\mnras] {10.1111/j.1365-2966.2007.11872.x}, \href {https://ui.adsabs.harvard.edu/abs/2007\mnras.378.1196K} {378, 1196}

\bibitem[\protect\citeauthoryear{{Kulkarni}, {Nichol}, {Sheth}, {Seo}, {Eisenstein}  \& {Gray}}{{Kulkarni} et~al.}{2007c}]{kulkarni2007three}
{Kulkarni} G.~V.,  {Nichol} R.~C.,  {Sheth} R.~K.,  {Seo} H.-J.,  {Eisenstein} D.~J.,   {Gray} A.,  2007c, \mn@doi [\mnras] {10.1111/j.1365-2966.2007.11872.x}, \href {https://ui.adsabs.harvard.edu/abs/2007MNRAS.378.1196K} {378, 1196}

\bibitem[\protect\citeauthoryear{Kulkarni, Nichol, Sheth, Seo, Eisenstein  \& Gray}{Kulkarni et~al.}{2007b}]{Kulkarni2007}
Kulkarni G.~V.,  Nichol R.~C.,  Sheth R.~K.,  Seo H.,  Eisenstein D.~J.,   Gray A.,  2007b, \mn@doi [\mnras] {10.1111/j.1365-2966.2007.11872.x}, 378, 1196

\bibitem[\protect\citeauthoryear{{Landy} \& {Szalay}}{{Landy} \& {Szalay}}{1993}]{LS1993}
{Landy} S.~D.,  {Szalay} A.~S.,  1993, \mn@doi [\apj] {10.1086/172900}, \href {https://ui.adsabs.harvard.edu/abs/1993\apj...412...64L} {412, 64}

\bibitem[\protect\citeauthoryear{{Laureijs} et~al.,}{{Laureijs} et~al.}{2011}]{Laureijs2011}
{Laureijs} R.,  et~al., 2011, \mn@doi [arXiv e-prints] {10.48550/arXiv.1110.3193}, \href {https://ui.adsabs.harvard.edu/abs/2011arXiv1110.3193L} {p. arXiv:1110.3193}

\bibitem[\protect\citeauthoryear{Li, Kauffmann, Jing, White, Börner  \& Cheng}{Li et~al.}{2006}]{Li2006}
Li C.,  Kauffmann G.,  Jing Y.~P.,  White S. D.~M.,  Börner G.,   Cheng F.~Z.,  2006, \mn@doi [\mnras] {10.1111/j.1365-2966.2006.10066.x}, 368, 21

\bibitem[\protect\citeauthoryear{Maartens, Fonseca, Camera, Jolicoeur, Viljoen  \& Clarkson}{Maartens et~al.}{2021}]{Maartens_2021}
Maartens R.,  Fonseca J.,  Camera S.,  Jolicoeur S.,  Viljoen J.-A.,   Clarkson C.,  2021, \mn@doi [\jcap] {10.1088/1475-7516/2021/12/009}, 2021, 009

\bibitem[\protect\citeauthoryear{{Manera} \& {Gazta{\~n}aga}}{{Manera} \& {Gazta{\~n}aga}}{2011}]{Manera2011}
{Manera} M.,  {Gazta{\~n}aga} E.,  2011, \mn@doi [\mnras] {10.1111/j.1365-2966.2011.18705.x}, \href {https://ui.adsabs.harvard.edu/abs/2011\mnras.415..383M} {415, 383}

\bibitem[\protect\citeauthoryear{March}{March}{2013}]{March2013}
March W.~B.,  2013, PhD thesis, Georgia Institute of Technology

\bibitem[\protect\citeauthoryear{{Mar{\'\i}n} et~al.,}{{Mar{\'\i}n} et~al.}{2013}]{MarinEtal2013}
{Mar{\'\i}n} F.~A.,  et~al., 2013, \mn@doi [\mnras] {10.1093/\mnras/stt520}, \href {https://ui.adsabs.harvard.edu/abs/2013\mnras.432.2654M} {432, 2654}

\bibitem[\protect\citeauthoryear{Marín}{Marín}{2011}]{Marín2011}
Marín F.,  2011, \mn@doi [\apj] {10.1088/0004-637x/737/2/97}, 737, 97

\bibitem[\protect\citeauthoryear{Marín, Wechsler, Frieman  \& Nichol}{Marín et~al.}{2008}]{Marín2008}
Marín F.~A.,  Wechsler R.~H.,  Frieman J.~A.,   Nichol R.~C.,  2008, \mn@doi [\apj] {10.1086/523628}, 672, 849

\bibitem[\protect\citeauthoryear{Massara, Villaescusa-Navarro, Ho, Dalal  \& Spergel}{Massara et~al.}{2021}]{Massara2021}
Massara E.,  Villaescusa-Navarro F.,  Ho S.,  Dalal N.,   Spergel D.~N.,  2021, \mn@doi [\prl] {10.1103/physrevlett.126.011301}, 126, 011301

\bibitem[\protect\citeauthoryear{{Matsubara}}{{Matsubara}}{2008}]{matsubara2008resumming}
{Matsubara} T.,  2008, \mn@doi [\prd] {10.1103/PhysRevD.77.063530}, \href {https://ui.adsabs.harvard.edu/abs/2008PhRvD..77f3530M} {77, 063530}

\bibitem[\protect\citeauthoryear{McBride, Connolly, Gardner, Scranton, Newman, Scoccimarro, Zehavi  \& Schneider}{McBride et~al.}{2010}]{mcbride2010three}
McBride C.~K.,  Connolly A.~J.,  Gardner J.~P.,  Scranton R.,  Newman J.~A.,  Scoccimarro R.,  Zehavi I.,   Schneider D.~P.,  2010, \mn@doi [\apj] {10.1088/0004-637X/726/1/13}, 726, 13

\bibitem[\protect\citeauthoryear{{McBride}, {Connolly}, {Gardner}, {Scranton}, {Newman}, {Scoccimarro}, {Zehavi}  \& {Schneider}}{{McBride} et~al.}{2011}]{McBrideEtal2011}
{McBride} C.~K.,  {Connolly} A.~J.,  {Gardner} J.~P.,  {Scranton} R.,  {Newman} J.~A.,  {Scoccimarro} R.,  {Zehavi} I.,   {Schneider} D.~P.,  2011, \mn@doi [\apj] {10.1088/0004-637X/726/1/13}, \href {https://ui.adsabs.harvard.edu/abs/2011\apj...726...13M} {726, 13}

\bibitem[\protect\citeauthoryear{{Meerburg} et~al.,}{{Meerburg} et~al.}{2019}]{Meerburg2019}
{Meerburg} P.~D.,  et~al., 2019, \mn@doi [\baas] {10.48550/arXiv.1903.04409}, \href {https://ui.adsabs.harvard.edu/abs/2019BAAS...51c.107M} {51, 107}

\bibitem[\protect\citeauthoryear{{Moore} et~al.,}{{Moore} et~al.}{2001}]{Moore2001}
{Moore} A.~W.,  et~al., 2001, in {Banday} A.~J.,  {Zaroubi} S.,   {Bartelmann} M.,  eds, Mining the Sky. p.~71 (\mn@eprint {arXiv} {astro-ph/0012333}), \mn@doi{10.1007/10849171_5}

\bibitem[\protect\citeauthoryear{Munshi, Melott  \& Coles}{Munshi et~al.}{2000}]{MunshiEtal2000}
Munshi D.,  Melott A.~L.,   Coles P.,  2000, \mn@doi [\mnras] {10.1046/j.1365-8711.2000.03042.x}, 311, 149

\bibitem[\protect\citeauthoryear{{Neyrinck}, {Szapudi}, {McCullagh}, {Szalay}, {Falck}  \& {Wang}}{{Neyrinck} et~al.}{2018}]{Neyrinck2018}
{Neyrinck} M.~C.,  {Szapudi} I.,  {McCullagh} N.,  {Szalay} A.~S.,  {Falck} B.,   {Wang} J.,  2018, \mn@doi [\mnras] {10.1093/\mnras/sty1074}, \href {https://ui.adsabs.harvard.edu/abs/2018\mnras.478.2495N} {478, 2495}

\bibitem[\protect\citeauthoryear{{Nichol} et~al.,}{{Nichol} et~al.}{2006}]{Nichol2006}
{Nichol} R.~C.,  et~al., 2006, \mn@doi [\mnras] {10.1111/j.1365-2966.2006.10239.x}, \href {https://ui.adsabs.harvard.edu/abs/2006\mnras.368.1507N} {368, 1507}

\bibitem[\protect\citeauthoryear{{Novell-Masot}, {Gualdi}, {Gil-Mar{\'\i}n}  \& {Verde}}{{Novell-Masot} et~al.}{2023}]{Novell-Masot2023}
{Novell-Masot} S.,  {Gualdi} D.,  {Gil-Mar{\'\i}n} H.,   {Verde} L.,  2023, \mn@doi [\jcap] {10.1088/1475-7516/2023/11/044}, \href {https://ui.adsabs.harvard.edu/abs/2023\jcap...11..044N} {2023, 044}

\bibitem[\protect\citeauthoryear{{Pan} \& {Szapudi}}{{Pan} \& {Szapudi}}{2005a}]{PanSzapudi2005a}
{Pan} J.,  {Szapudi} I.,  2005a, \mn@doi [\mnras] {10.1111/j.1365-2966.2005.09177.x}, \href {https://ui.adsabs.harvard.edu/abs/2005\mnras.361..357P} {361, 357}

\bibitem[\protect\citeauthoryear{Pan \& Szapudi}{Pan \& Szapudi}{2005b}]{PanSzapudi2005b}
Pan J.,  Szapudi I.,  2005b, \mn@doi [\mnras] {10.1111/j.1365-2966.2005.09407.x}, 362, 1363

\bibitem[\protect\citeauthoryear{Pando \& Fang}{Pando \& Fang}{1998}]{Fang1998Book}
Pando J.,  Fang L.-Z.,  1998, An Introduction of The Wavelet Transform.
Singapore: World Scientific, pp 15--45

\bibitem[\protect\citeauthoryear{{Pearson} \& {Samushia}}{{Pearson} \& {Samushia}}{2018}]{Pearson2018}
{Pearson} D.~W.,  {Samushia} L.,  2018, \mn@doi [\mnras] {10.1093/\mnras/sty1266}, \href {https://ui.adsabs.harvard.edu/abs/2018\mnras.478.4500P} {478, 4500}

\bibitem[\protect\citeauthoryear{Peebles}{Peebles}{1980}]{peebles1980}
Peebles P.,  1980, The Large Scale Structure of the Universe.
Princeton: Princeton Univ. Press

\bibitem[\protect\citeauthoryear{{Peebles} \& {Groth}}{{Peebles} \& {Groth}}{1975}]{Peebles1975}
{Peebles} P.~J.~E.,  {Groth} E.~J.,  1975, \mn@doi [\apj] {10.1086/153390}, \href {https://ui.adsabs.harvard.edu/abs/1975ApJ...196....1P} {196, 1}

\bibitem[\protect\citeauthoryear{Percival et~al.,}{Percival et~al.}{2010}]{Percival2011}
Percival W.~J.,  et~al., 2010, \mn@doi [\mnras] {10.1111/j.1365-2966.2009.15812.x}, 401, 2148

\bibitem[\protect\citeauthoryear{{Percival} et~al.,}{{Percival} et~al.}{2014}]{PercivalEtal2014}
{Percival} W.~J.,  et~al., 2014, \mn@doi [\mnras] {10.1093/\mnras/stu112}, \href {https://ui.adsabs.harvard.edu/abs/2014\mnras.439.2531P} {439, 2531}

\bibitem[\protect\citeauthoryear{Philcox \& Slepian}{Philcox \& Slepian}{2022}]{PhilcoxSlepian2022}
Philcox O. H.~E.,  Slepian Z.,  2022, \mn@doi [Proceedings of the National Academy of Sciences] {10.1073/pnas.2111366119}, 119, e2111366119

\bibitem[\protect\citeauthoryear{Philcox, Massara  \& Spergel}{Philcox et~al.}{2020}]{Philcox2020}
Philcox O. H.~E.,  Massara E.,   Spergel D.~N.,  2020, \mn@doi [\prd] {10.1103/physrevd.102.043516}, 102, 043516

\bibitem[\protect\citeauthoryear{{Philcox}, {Slepian}, {Hou}, {Warner}, {Cahn}  \& {Eisenstein}}{{Philcox} et~al.}{2022}]{PhilcoxEtal2022}
{Philcox} O. H.~E.,  {Slepian} Z.,  {Hou} J.,  {Warner} C.,  {Cahn} R.~N.,   {Eisenstein} D.~J.,  2022, \mn@doi [\mnras] {10.1093/\mnras/stab3025}, \href {https://ui.adsabs.harvard.edu/abs/2022\mnras.509.2457P} {509, 2457}

\bibitem[\protect\citeauthoryear{{Prada}, {Klypin}, {Cuesta}, {Betancort-Rijo}  \& {Primack}}{{Prada} et~al.}{2012}]{prada2012halo}
{Prada} F.,  {Klypin} A.~A.,  {Cuesta} A.~J.,  {Betancort-Rijo} J.~E.,   {Primack} J.,  2012, \mn@doi [\mnras] {10.1111/j.1365-2966.2012.21007.x}, \href {https://ui.adsabs.harvard.edu/abs/2012MNRAS.423.3018P} {423, 3018}

\bibitem[\protect\citeauthoryear{{Pujol}, {Hoffmann}, {Jim{\'e}nez}  \& {Gazta{\~n}aga}}{{Pujol} et~al.}{2017}]{Pujol2017}
{Pujol} A.,  {Hoffmann} K.,  {Jim{\'e}nez} N.,   {Gazta{\~n}aga} E.,  2017, \mn@doi [\aap] {10.1051/0004-6361/201629121}, \href {https://ui.adsabs.harvard.edu/abs/2017A&A...598A.103P} {598, A103}

\bibitem[\protect\citeauthoryear{{Riebe} et~al.,}{{Riebe} et~al.}{2011}]{riebe2013multidark}
{Riebe} K.,  et~al., 2011, \mn@doi [arXiv e-prints] {10.48550/arXiv.1109.0003}, \href {https://ui.adsabs.harvard.edu/abs/2011arXiv1109.0003R} {p. arXiv:1109.0003}

\bibitem[\protect\citeauthoryear{Saga, Taruya, Rasera  \& Breton}{Saga et~al.}{2022}]{Saga2022}
Saga S.,  Taruya A.,  Rasera Y.,   Breton M.-A.,  2022, \mn@doi [\mnras] {10.1093/\mnras/stac186}, 511, 2732

\bibitem[\protect\citeauthoryear{{Samushia}, {Slepian}  \& {Villaescusa-Navarro}}{{Samushia} et~al.}{2021}]{Samushia2021}
{Samushia} L.,  {Slepian} Z.,   {Villaescusa-Navarro} F.,  2021, \mn@doi [\mnras] {10.1093/\mnras/stab1199}, \href {https://ui.adsabs.harvard.edu/abs/2021\mnras.505..628S} {505, 628}

\bibitem[\protect\citeauthoryear{{S{\'a}nchez}, {Baugh}  \& {Angulo}}{{S{\'a}nchez} et~al.}{2008}]{Sanchez2008}
{S{\'a}nchez} A.~G.,  {Baugh} C.~M.,   {Angulo} R.~E.,  2008, \mn@doi [\mnras] {10.1111/j.1365-2966.2008.13769.x}, \href {https://ui.adsabs.harvard.edu/abs/2008MNRAS.390.1470S} {390, 1470}

\bibitem[\protect\citeauthoryear{Satpathy, A C Croft, Ho  \& Li}{Satpathy et~al.}{2019}]{Satpathy2019}
Satpathy S.,  A C Croft R.,  Ho S.,   Li B.,  2019, \mn@doi [\mnras] {10.1093/\mnras/stz009}, 484, 2148

\bibitem[\protect\citeauthoryear{{Schulz}}{{Schulz}}{2023}]{Schulz2023}
{Schulz} S.,  2023, \mn@doi [\mnras] {10.1093/\mnras/stad2868}, \href {https://ui.adsabs.harvard.edu/abs/2023\mnras.526.3951S} {526, 3951}

\bibitem[\protect\citeauthoryear{{Scoccimarro}, {Feldman}, {Fry}  \& {Frieman}}{{Scoccimarro} et~al.}{2001}]{Scoccimarro2001}
{Scoccimarro} R.,  {Feldman} H.~A.,  {Fry} J.~N.,   {Frieman} J.~A.,  2001, \mn@doi [\apj] {10.1086/318284}, \href {https://ui.adsabs.harvard.edu/abs/2001\apj...546..652S} {546, 652}

\bibitem[\protect\citeauthoryear{{Scoccimarro}, {Hui}, {Manera}  \& {Chan}}{{Scoccimarro} et~al.}{2012}]{Scoccimarro2012}
{Scoccimarro} R.,  {Hui} L.,  {Manera} M.,   {Chan} K.~C.,  2012, \mn@doi [\prd] {10.1103/PhysRevD.85.083002}, \href {https://ui.adsabs.harvard.edu/abs/2012PhRvD..85h3002S} {85, 083002}

\bibitem[\protect\citeauthoryear{Sheth}{Sheth}{2005}]{Sheth2005}
Sheth R.~K.,  2005, \mn@doi [\mnras] {10.1111/j.1365-2966.2005.09609.x}, 364, 796

\bibitem[\protect\citeauthoryear{Sheth \& Tormen}{Sheth \& Tormen}{2004}]{Sheth2004}
Sheth R.~K.,  Tormen G.,  2004, \mn@doi [\mnras] {10.1111/j.1365-2966.2004.07733.x}, 350, 1385

\bibitem[\protect\citeauthoryear{{Simpson}, {James}, {Heavens}  \& {Heymans}}{{Simpson} et~al.}{2011}]{Simpson2011}
{Simpson} F.,  {James} J.~B.,  {Heavens} A.~F.,   {Heymans} C.,  2011, \mn@doi [\prl] {10.1103/PhysRevLett.107.271301}, \href {https://ui.adsabs.harvard.edu/abs/2011PhRvL.107A1301S} {107, 271301}

\bibitem[\protect\citeauthoryear{{Simpson}, {Heavens}  \& {Heymans}}{{Simpson} et~al.}{2013}]{Simpson2013}
{Simpson} F.,  {Heavens} A.~F.,   {Heymans} C.,  2013, \mn@doi [\prd] {10.1103/PhysRevD.88.083510}, \href {https://ui.adsabs.harvard.edu/abs/2013PhRvD..88h3510S} {88, 083510}

\bibitem[\protect\citeauthoryear{{Singh}}{{Singh}}{2021}]{Singh2021}
{Singh} S.,  2021, \mn@doi [\mnras] {10.48550/arXiv.2105.04548}, \href {https://ui.adsabs.harvard.edu/abs/2021\mnras.508.1632S} {508, 1632}

\bibitem[\protect\citeauthoryear{{Sinha} \& {Garrison}}{{Sinha} \& {Garrison}}{2020}]{Sinha2020}
{Sinha} M.,  {Garrison} L.~H.,  2020, \mn@doi [\mnras] {10.1093/\mnras/stz3157}, \href {https://ui.adsabs.harvard.edu/abs/2020\mnras.491.3022S} {491, 3022}

\bibitem[\protect\citeauthoryear{Skibba, Sheth, Connolly  \& Scranton}{Skibba et~al.}{2006}]{Skibba2006}
Skibba R.,  Sheth R.~K.,  Connolly A.~J.,   Scranton R.,  2006, \mn@doi [\mnras] {10.1111/j.1365-2966.2006.10196.x}, 369, 68

\bibitem[\protect\citeauthoryear{Skibba, Sheth, Croton, Muldrew, Abbas, Pearce  \& Shattow}{Skibba et~al.}{2013}]{Skibba2013}
Skibba R.~A.,  Sheth R.~K.,  Croton D.~J.,  Muldrew S.~I.,  Abbas U.,  Pearce F.~R.,   Shattow G.~M.,  2013, \mn@doi [\mnras] {10.1093/\mnras/sts349}, 429, 458

\bibitem[\protect\citeauthoryear{Slepian \& Eisenstein}{Slepian \& Eisenstein}{2015}]{SlepianEisenstein2016}
Slepian Z.,  Eisenstein D.~J.,  2015, \mn@doi [Monthly Notices of the Royal Astronomical Society: Letters] {10.1093/mnrasl/slv133}, 455, L31

\bibitem[\protect\citeauthoryear{Slepian \& Eisenstein}{Slepian \& Eisenstein}{2018}]{Slepian2018}
Slepian Z.,  Eisenstein D.~J.,  2018, \mn@doi [\mnras] {10.1093/\mnras/sty1063}, 478, 1468

\bibitem[\protect\citeauthoryear{{Slepian} et~al.,}{{Slepian} et~al.}{2017}]{Slepian2017}
{Slepian} Z.,  et~al., 2017, \mn@doi [\mnras] {10.1093/\mnras/stx488}, \href {https://ui.adsabs.harvard.edu/abs/2017\mnras.469.1738S} {469, 1738}

\bibitem[\protect\citeauthoryear{Smith}{Smith}{2009}]{Smith2009}
Smith R.~E.,  2009, \mn@doi [\mnras] {10.1111/j.1365-2966.2009.15490.x}, 400, 851

\bibitem[\protect\citeauthoryear{Smith et~al.,}{Smith et~al.}{2003}]{smith2003stable}
Smith R.~E.,  et~al., 2003, Monthly Notices of the Royal Astronomical Society, 341, 1311

\bibitem[\protect\citeauthoryear{{Sosa Nu{\~n}ez} \& {Niz}}{{Sosa Nu{\~n}ez} \& {Niz}}{2020}]{SosaNiz2020}
{Sosa Nu{\~n}ez} F.,  {Niz} G.,  2020, \mn@doi [\jcap] {10.1088/1475-7516/2020/12/021}, \href {https://ui.adsabs.harvard.edu/abs/2020JCAP...12..021S} {2020, 021}

\bibitem[\protect\citeauthoryear{{Storey-Fisher} \& {Hogg}}{{Storey-Fisher} \& {Hogg}}{2021}]{Fisher2021}
{Storey-Fisher} K.,  {Hogg} D.~W.,  2021, \mn@doi [\apj] {10.3847/1538-4357/abdc21}, \href {https://ui.adsabs.harvard.edu/abs/2021\apj...909..220S} {909, 220}

\bibitem[\protect\citeauthoryear{{Sugiyama} et~al.,}{{Sugiyama} et~al.}{2023}]{Sugiyama2023}
{Sugiyama} N.~S.,  et~al., 2023, \mn@doi [\mnras] {10.1093/\mnras/stad1505}, \href {https://ui.adsabs.harvard.edu/abs/2023\mnras.523.3133S} {523, 3133}

\bibitem[\protect\citeauthoryear{{Sunseri}, {Slepian}, {Portillo}, {Hou}, {Kahraman}  \& {Finkbeiner}}{{Sunseri} et~al.}{2023}]{Sunseri2023}
{Sunseri} J.,  {Slepian} Z.,  {Portillo} S.,  {Hou} J.,  {Kahraman} S.,   {Finkbeiner} D.~P.,  2023, \mn@doi [RAS Techniques and Instruments] {10.1093/rasti/rzad003}, \href {https://ui.adsabs.harvard.edu/abs/2023RASTI...2...62S} {2, 62}

\bibitem[\protect\citeauthoryear{{Szapudi}}{{Szapudi}}{1998}]{Szapudi1998}
{Szapudi} I.,  1998, \mn@doi [\mnras] {10.1046/j.1365-8711.1998.300004l35.x}, \href {https://ui.adsabs.harvard.edu/abs/1998\mnras.300L..35S} {300, L35}

\bibitem[\protect\citeauthoryear{{Szapudi}}{{Szapudi}}{2004}]{Szapudi2004}
{Szapudi} I.,  2004, \mn@doi [\apjl] {10.1086/420894}, \href {https://ui.adsabs.harvard.edu/abs/2004\apj...605L..89S} {605, L89}

\bibitem[\protect\citeauthoryear{Szapudi}{Szapudi}{2009}]{Szapudi2009}
Szapudi I.,  2009, Introduction to Higher Order Spatial Statistics in Cosmology.
Springer Berlin Heidelberg, Berlin, Heidelberg, pp 457--492, \mn@doi{10.1007/978-3-540-44767-2_14}, \url {https://doi.org/10.1007/978-3-540-44767-2_14}

\bibitem[\protect\citeauthoryear{{Szapudi} \& {Colombi}}{{Szapudi} \& {Colombi}}{1996}]{SzapudiColombi1996}
{Szapudi} I.,  {Colombi} S.,  1996, \mn@doi [\apj] {10.1086/177855}, \href {https://ui.adsabs.harvard.edu/abs/1996\apj...470..131S} {470, 131}

\bibitem[\protect\citeauthoryear{{Szapudi} \& {Szalay}}{{Szapudi} \& {Szalay}}{1997}]{SzapudiSzalay1997}
{Szapudi} I.,  {Szalay} A.~S.,  1997, \mn@doi [\apjl] {10.1086/310641}, \href {https://ui.adsabs.harvard.edu/abs/1997\apj...481L...1S} {481, L1}

\bibitem[\protect\citeauthoryear{{Szapudi} \& {Szalay}}{{Szapudi} \& {Szalay}}{1998}]{SzapudiSzalay1998}
{Szapudi} I.,  {Szalay} A.~S.,  1998, \mn@doi [\apjl] {10.1086/311146}, \href {https://ui.adsabs.harvard.edu/abs/1998\apj...494L..41S} {494, L41}

\bibitem[\protect\citeauthoryear{Tansella, Bonvin, Durrer, Ghosh  \& Sellentin}{Tansella et~al.}{2018}]{Tansella_2018}
Tansella V.,  Bonvin C.,  Durrer R.,  Ghosh B.,   Sellentin E.,  2018, \mn@doi [\jcap] {10.1088/1475-7516/2018/03/019}, 2018, 019

\bibitem[\protect\citeauthoryear{{Tegmark} et~al.,}{{Tegmark} et~al.}{2006}]{Tegmark2006}
{Tegmark} M.,  et~al., 2006, \mn@doi [\prd] {10.1103/PhysRevD.74.123507}, \href {https://ui.adsabs.harvard.edu/abs/2006PhRvD..74l3507T} {74, 123507}

\bibitem[\protect\citeauthoryear{Tessore}{Tessore}{2018}]{Tessore2018}
Tessore N.,  2018, \mn@doi [Research Notes of the AAS] {10.3847/2515-5172/aad9a7}, 2, 148

\bibitem[\protect\citeauthoryear{Valogiannis \& Bean}{Valogiannis \& Bean}{2018}]{Valogiannis2018}
Valogiannis G.,  Bean R.,  2018, \mn@doi [\prd] {10.1103/physrevd.97.023535}, 97, 023535

\bibitem[\protect\citeauthoryear{Vargas-Maga{\~n}a et~al.,}{Vargas-Maga{\~n}a et~al.}{2013}]{vargas2013optimized}
Vargas-Maga{\~n}a M.,  et~al., 2013, \aap, 554, A131

\bibitem[\protect\citeauthoryear{{Veropalumbo} et~al.,}{{Veropalumbo} et~al.}{2021}]{Veropalumbo2021}
{Veropalumbo} A.,  et~al., 2021, \mn@doi [\mnras] {10.1093/\mnras/stab2205}, \href {https://ui.adsabs.harvard.edu/abs/2021\mnras.507.1184V} {507, 1184}

\bibitem[\protect\citeauthoryear{{Villaescusa-Navarro} et~al.,}{{Villaescusa-Navarro} et~al.}{2020}]{villaescusa2020quijote}
{Villaescusa-Navarro} F.,  et~al., 2020, \mn@doi [\apjs] {10.3847/1538-4365/ab9d82}, \href {https://ui.adsabs.harvard.edu/abs/2020ApJS..250....2V} {250, 2}

\bibitem[\protect\citeauthoryear{Wechsler \& Tinker}{Wechsler \& Tinker}{2018}]{Wechsler2018}
Wechsler R.~H.,  Tinker J.~L.,  2018, \mn@doi [Annual Review of Astronomy and Astrophysics] {10.1146/annurev-astro-081817-051756}, 56, 435

\bibitem[\protect\citeauthoryear{White}{White}{2016}]{White2016}
White M.,  2016, \mn@doi [\jcap] {10.1088/1475-7516/2016/11/057}, 2016, 057

\bibitem[\protect\citeauthoryear{White \& Padmanabhan}{White \& Padmanabhan}{2009}]{White2009}
White M.,  Padmanabhan N.,  2009, \mn@doi [\mnras] {10.1111/j.1365-2966.2009.14732.x}, 395, 2381

\bibitem[\protect\citeauthoryear{Xiao et~al.,}{Xiao et~al.}{2022}]{Xiao2022}
Xiao X.,  et~al., 2022, \mn@doi [\mnras] {10.1093/\mnras/stac879}, 513, 595

\bibitem[\protect\citeauthoryear{{Xu}, {Padmanabhan}, {Eisenstein}, {Mehta}  \& {Cuesta}}{{Xu} et~al.}{2012}]{Xu2012}
{Xu} X.,  {Padmanabhan} N.,  {Eisenstein} D.~J.,  {Mehta} K.~T.,   {Cuesta} A.~J.,  2012, \mn@doi [\mnras] {10.1111/j.1365-2966.2012.21573.x}, \href {https://ui.adsabs.harvard.edu/abs/2012\mnras.427.2146X} {427, 2146}

\bibitem[\protect\citeauthoryear{Yang}{Yang}{2010}]{Yang2010}
Yang Y.-F.,  2010, \mn@doi [Chinese Astronomy and Astrophysics] {10.1016/j.chinastron.2010.07.010}, 34, 255

\bibitem[\protect\citeauthoryear{Yang, Feng, Chu  \& Fang}{Yang et~al.}{2001a}]{yang2001a}
Yang X.,  Feng L.-L.,  Chu Y.,   Fang L.-Z.,  2001a, \mn@doi [\apj] {10.1086/320661}, 553, 1

\bibitem[\protect\citeauthoryear{Yang, Feng, Chu  \& Fang}{Yang et~al.}{2001b}]{yang2001b}
Yang X.~H.,  Feng L.-L.,  Chu Y.~Q.,   Fang L.-Z.,  2001b, \mn@doi [\apj] {10.1086/323059}, 560, 549

\bibitem[\protect\citeauthoryear{Yang, Feng, Chu  \& Fang}{Yang et~al.}{2002}]{yang2002}
Yang X.,  Feng L.-L.,  Chu Y.,   Fang L.-Z.,  2002, \mn@doi [\apj] {10.1086/338274}, 566, 630

\bibitem[\protect\citeauthoryear{Yang, Mo  \& Bosch}{Yang et~al.}{2003}]{Yang2003}
Yang X.,  Mo H.~J.,   Bosch F. C. v.~d.,  2003, \mn@doi [\mnras] {10.1046/j.1365-8711.2003.06254.x}, 339, 1057

\bibitem[\protect\citeauthoryear{{Zehavi} et~al.,}{{Zehavi} et~al.}{2011}]{zehavi2011}
{Zehavi} I.,  et~al., 2011, \mn@doi [\apj] {10.1088/0004-637X/736/1/59}, \href {https://ui.adsabs.harvard.edu/abs/2011ApJ...736...59Z} {736, 59}

\bibitem[\protect\citeauthoryear{{Zhan}}{{Zhan}}{2011}]{Zhan2011}
{Zhan} H.,  2011, \mn@doi [Scientia Sinica Physica, Mechanica \& Astronomica] {10.1360/132011-961}, \href {https://ui.adsabs.harvard.edu/abs/2011SSPMA..41.1441Z} {41, 1441}

\bibitem[\protect\citeauthoryear{{Zhao}}{{Zhao}}{2023}]{Zhao2023}
{Zhao} C.,  2023, \mn@doi [\aap] {10.1051/0004-6361/202346015}, \href {https://ui.adsabs.harvard.edu/abs/2023A&A...672A..83Z} {672, A83}

\makeatother
\end{thebibliography}

%%%%%%%%%%%%%%%%%%%%%%%%%%%%%%%%%%%%%%%%%%%%%%%%%%

%%%%%%%%%%%%%%%%% APPENDICES %%%%%%%%%%%%%%%%%%%%%

\appendix

\section{3PCF: convergence and performance tests}

\begin{figure*}
    \centering
    \includegraphics[width=\textwidth]{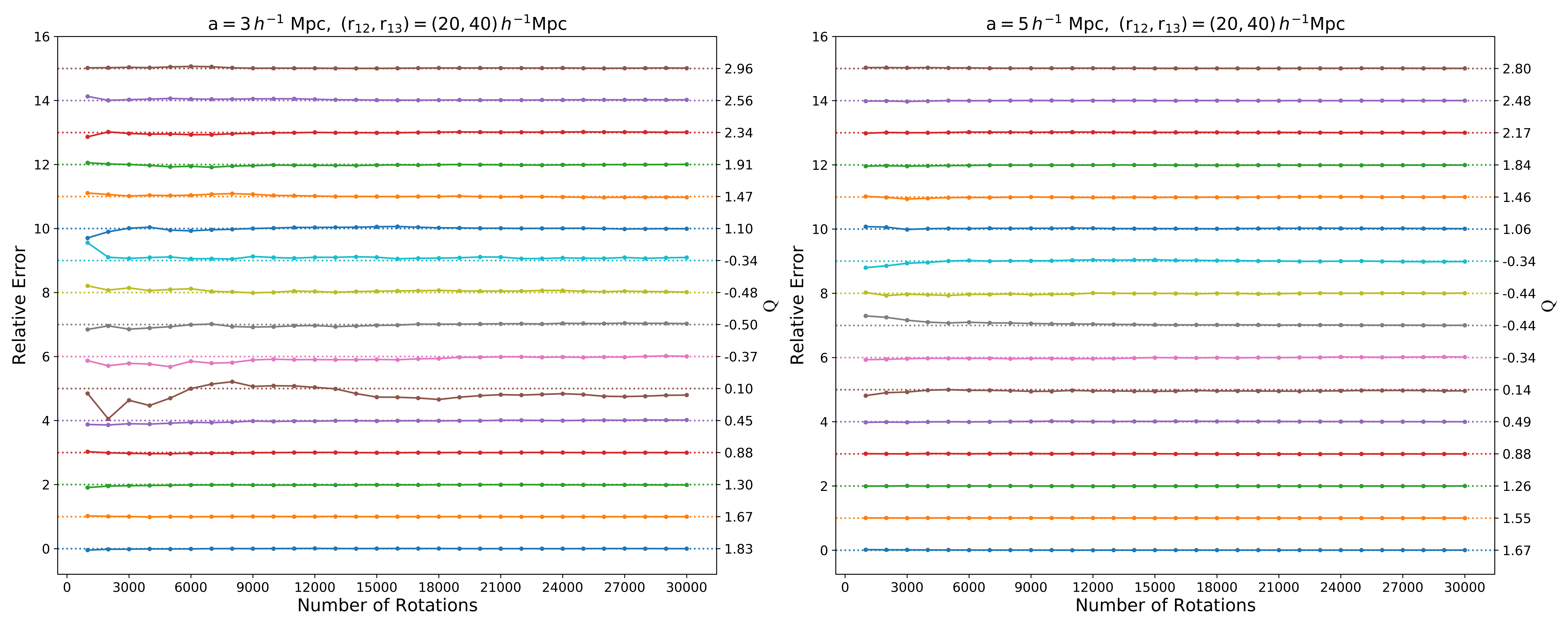}
    \caption{The convergence test of the Q factor. We plotted relative errors of $Q(r_{12},r_{13},\cos\theta)$ for the configuration $(r_{12},r_{13}) = (20,60)h^{-1}$Mpc at various angles $\theta= \arccos(\hat{\bf r}_{12}\cdot\hat{\bf r}_{13})$ with an increasing number of random spatial orientations. For clarity, the relative error of Q is offset vertically by one unit in turn with the angle $\theta$ increasing from $0$ (lower) to $\pi$ (upper), and the corresponding Q values are marked on the right y-axis. We also compare the relative error for the different binning schemes with $a=3h^{-1}$Mpc (left panel) and $a=5h^{-1}$Mpc (right panel).}
    \label{fig:convergence-test-2040}
\end{figure*}

As discussed in Section \ref{sec:3PCF-binning}, one triplet counting scheme in computing the 3PCF is to assign the primary vertex of a triangle to each data point and then rotate around it to obtain an average over spatial orientations. For each point, the rotation number is roughly estimated to be $N_{\text{rot}}\sim (nV_{R_{\text {max}}})^2$, where $n$ is the number density of the catalogue, and $V_{R_{\text{max}}}$ is the volume of the sphere of radius $R_{\text max}$, where $R_{\text{max}}$ is the maximum scale to be measured in 3PCF \citep{March2013}. For the Quijote halo catalogues, the number density is around $4.0\times 10^{-4}(h^{-1}\text{Mpc})^{-3}$, and thus 
\begin{equation}\label{eq:3PCF-num.rotation}
N_{\text{rot}} \sim 4\times 10^4 \Bigl(\frac{n}{4.0\times10^{-4}(h^{-1}\text{Mpc})^{-3}}\Bigr)^2\Bigl(\frac{R_{\text{max}}}{50h^{-1}\text{Mpc}}\Bigr)^6  
\end{equation}
The equation above estimates the number of relevant triangles for each data point, which has been shown to be a reliable choice for achieving convergence in 3PCF measurements for a given triangle shape. However, within a specific binning scheme, the question arises of whether there is an optimal or minimum sampling number that ensures convergent results in a Monte Carlo experiment, or rather, whether the convergence rate is influenced by the binning scheme. Similarly to the binning effect in 2PCF measurements, a large bin size can effectively suppress shot noise on small scales, but this comes at the cost of information loss due to the smearing of the shape dependence of the $Q$ factor. In this study, we will focus on convergence and performance tests for 3PCF measurements in the Quijote halo catalogues using different binning schemes, evaluating how binning affects both the convergence rate and computational performance. 

In this paper, we consider only two triangle configurations in the 3PCF measurement, $(r_{12},r_{13}) = (20,40)h^{-1}$Mpc and $(20,60)h^{-1}$Mpc. For the binning scheme, we adopted the filtered radius of 3, 5, 8 $h^{-1}$Mpc, among which the largest value 8$h^{-1}$Mpc is $7.5\%$ and $10.0\%$ of the maximum scale explored for the configuration $(r_{12},r_{13}) = (20,40), \, (20,60)h^{-1}$Mpc, respectively. As demonstrated in section \ref{sec:3PCF_test}, this bin size does not produce a significant impact on the shape dependence of 3PCF.

\begin{figure*}
    \centering
    \includegraphics[width=\textwidth]{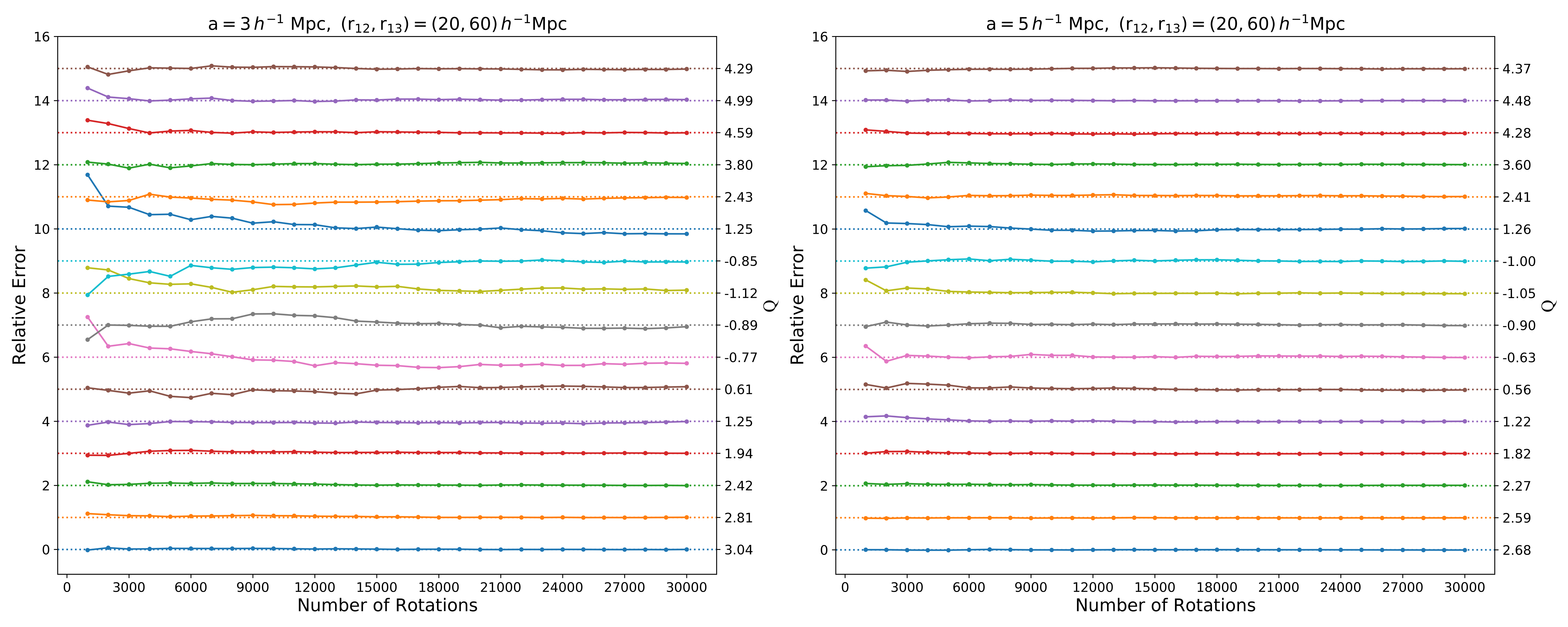}
    \caption{The same as Fig.~\ref{fig:convergence-test-2040} but for the configuration $(r_{12},r_{13}) = (20,60)h^{-1}$Mpc.}
    \label{fig:convergence-test-2060}
\end{figure*}

For an $N_p$-object catalogue, there are $N_pN_{\text{rot}}$ relevant triangles produced by applying $N_{\text{rot}}$ rotations around each point and then traversing all $N_p$ points. By averaging over $N=N_pN_{\text{rot}}$ triangles, we compute the $Q$-factor, denoted by $Q_N$, incrementing the number of rotations by 1,000 at each step until reaching a predefined maximum $N_{\text{rot,max}}$. According to Eq.(\ref{eq:3PCF-num.rotation}), we set $N_{\text{rot,max}}=10^5$ as a safe upper limit for the two triangle configurations considered in this paper. We take the final output as the convergence value of the $Q$-factor, denoted as $Q_c$ and go back to evaluate the relative error for each output, defined by $\Delta_Q = Q_N/Q_c -1$. The results are shown in Fig.~\ref{fig:convergence-test-2040} and Fig.~\ref{fig:convergence-test-2060} for the two triangle configurations, respectively. 

Figure~\ref{fig:convergence-test-2040} shows the relative errors $\Delta_Q$ in the configuration $(r_{12},r_{13}) = (20,40) h^{-1}$Mpc as the number of random spatial orientations increases, with bin sizes of 3 $h^{-1}$Mpc (left panel) and 5 $h^{-1}$Mpc (right panel). In the former case, the bin size is slightly smaller than the spatial resolution set by the basis function with $J=8$, corresponding to a grid size of 4.0 $h^{-1}$Mpc. It is evident that, except for small values $Q \lesssim 0.5$, $Q$ reaches the convergence value with a relative error of $\lesssim 5.0\%$ at $N_{\text{rot}} \lesssim 10^4$. In particular, for nearly elongated configurations, convergence can be achieved around $N_{\text{rot}}\sim 5000$. When using a slightly larger bin size of $a=5 h^{-1}$Mpc, as shown in the right panel of Fig.~\ref{fig:convergence-test-2040}, the convergence rate improves significantly. For most $Q$ values, the number of random spatial orientations required to achieve a relative error of $\sim 2.0\%$ drops to $N_{\text{rot}}=2000$. Moreover, in a numerical test with $a=8 h^{-1}$Mpc, the minimum number of rotations needed for convergence is further reduced to $N_{\text{rot}} = 500$, indicating much faster convergence compared to bin sizes of $a=3$ or $5 h^{-1}$Mpc.

Similarly, Fig.\ref{fig:convergence-test-2060} presents the convergence test for triangle configurations with side lengths expanded to $(r_{12}, r_{13}) = (20, 60), h^{-1}$Mpc, using the same binning scheme as in Fig.\ref{fig:convergence-test-2040}. This binning scheme decreases the ratio of bin size to the maximum probed scale, leading to increased shot noise due to fewer triplet counts within bins, as clustering weakens on larger scales. Consequently, the convergence is noticeably weaker than in Fig.~\ref{fig:convergence-test-2040} for a given number of rotations, especially at rectangular configurations. However, as expected, convergence can be significantly refined with an increased filter radius. Hence, to improve both the convergence rate and computational efficiency, it is strongly suggested that a scale-adaptive binning scheme be adopted in practical 3PCF measurements. Nonetheless, determining an optimal filtered radius remains an issue that needs to be addressed in future studies. 

In addition, we also analysed the convergence of the average Q factor (see Fig.~\ref{fig:quijote-halo-3pcf}) in the Quijote 50 halo samples as a function of the number of rotations. The result indicates that even in the worst convergence case of $(r_{12},r_{13}) = (20,60)h^{-1}$Mpc with a filtered radius $3h^{-1}$Mpc, the averaged Q factor can reach its stable value quickly as early as $N_{\text{rot}} = 1000$, and is barely impacted by further increases in the number of rotations. This result suggests that the primary factor affecting convergence is Poisson shot noise, and the stacking of independent samples can significantly reduce this effect.   

\begin{figure}
    \centering
    \includegraphics[width=8.0cm]{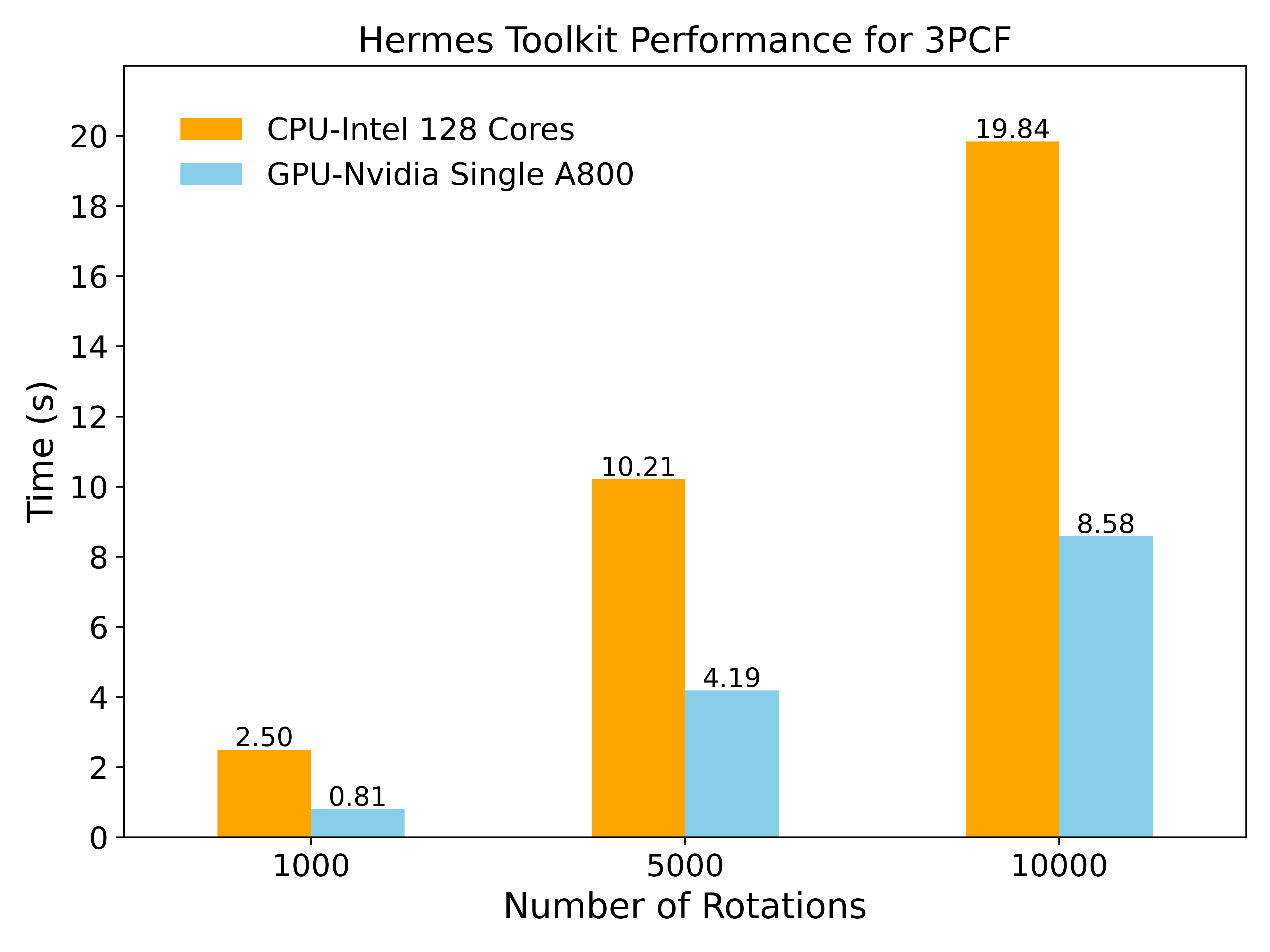}
    \caption{Performance tests: CPU time in units of second for measuring the Q factor at one angle with increasing number of rotations $N_{\text{rot}}$. We also compare the performance of the {\tt Hermes} toolkit between an Intel 128Cores server and a single Nvidia GPU A800 card.}
    \label{fig:hardware-test}
\end{figure}

In this work, we perform 2/3PCF computations using two distinct working versions of the {\tt Hermes} toolkit, which were updated from the original {\tt MRACS} code written in Fortran 90. One of the revised versions is a C++ implementation with OpenMP parallelisation, and the other is an MPI parallelised, GPU-accelerated package written in Python. Runtime performance tests were conducted on two hardware systems: (1) a compute server equipped with 128 cores of Intel(R) Xeon(R) Gold 5318H CPUs @ 2.50GHz and 1TB shared memory, and (2) a single NVIDIA A800 PCIe 80GB card within an 8-GPU stacked node in the Tianhe supercomputer (Guangzhou).    

Fig.~\ref{fig:hardware-test} compares the task performance between two hardware systems, showing the runtime for three typical rotation numbers in the 3PCF calculation. The 3PCF calculation in {\tt Hermes} is based on triplet-counting within 3-spheres. Using the \cite{SzapudiSzalay1998} estimator, a measurement with $N_{\text{rot}}=10^4$ rotations in a Quijote halo catalogue requires approximately $2.4\times 10^{10}$ CIC operations. As illustrated in Fig.~\ref{fig:hardware-test}, this computation takes 19.84 seconds on the CPU machine and 8.58 seconds on the GPU card, making the GPU only 2.3 times faster than the CPU. A detailed run-time analysis reveals that on the A800 card, the $80\%$ execution time is spent on allocating data to the cache of each streaming multiprocessor, with only $20\%$ of the time dedicated to floating-point computations. Taking into account this factor, the FP64 performance of the A800 GPU should technically offer nearly ten times the acceleration over the CPU system, which is in agreement with industry benchmarks for these two systems. 

Fig.~\ref{fig:hardware-test} also shows an approximately linear relationship between runtime and task load. If the number of orientations decreases to $N=2000$, which is sufficiently large to ensure convergence at the filtered radius of $a=5 $h$^{-1}$ Mpc, the entire $Q$ curve evaluated at 20 sampling angles, as shown in Fig.~\ref{fig:quijote-halo-3pcf}, can be obtained in roughly 40 seconds. 

In short, the {\tt Hermes} algorithm demonstrates highly efficient parallelisation in multithreaded CIC computations. Benchmarking shows that timing scales linearly with the number of threads or triplets, manifesting the scalability and efficiency of the algorithm. 

%%%%%%%%%%%%%%%%%%%%%%%%%%%%%%%%%%%%%%%%%%%%%%%%%%

% Don't change these lines
\bsp	% typesetting comment
\label{lastpage}
\end{document}